\documentstyle[prc,aps,epsf]{revtex}

\begin{document}
\draft

\title{\hskip 6.0truein BNL-62600\\ \hfill \\ \hfill \\
Modeling Cluster Production at the AGS}  


\author{D.~E.~Kahana$^{2}$, S.~H.~Kahana$^{1}$, Y.~Pang$^{1,3}$, \\
  A.~J.~Baltz$^{1}$, C.~B.~Dover$^{1}$, E.~Schnedermann$^{1}$, \\
   T.~J.~Schlagel$^{1,2}$}
\address{$^{1}$Physics Department, Brookhaven National Laboratory\\
   Upton, NY 11973, USA\\
   $^{2}$Physics Department, State University of New York, \\
   Stony Brook, NY 11791, USA\\
   $^{3}$Physics Department, Columbia University, \\
   New York, NY 10027, USA}  
\date{\today}  
  
\maketitle  
  
\begin{abstract}  
Deuteron coalescence, during relativistic nucleus-nucleus collisions, is
carried out in a model incorporating a minimal quantal treatment of the
formation of the cluster from its individual nucleons by evaluating the
overlap of intial cascading nucleon wave packets with the final deuteron wave
function. In one approach the nucleon and deuteron center of mass wave packet
sizes are estimated dynamically for each coalescing pair using its past
light-cone history in the underlying cascade, a procedure which yields a
parameter free determination of the cluster yield.  A modified version
employing a global estimate of the deuteron formation probability, is
identical to a general implementation of the Wigner function formalism but
can differ from the most frequent realisation of the latter. Comparison is
made both with the extensive existing E802 data for Si+Au at 14.6 GeV/c and
with the Wigner formalism.  A globally consistent picture of the Si+Au
measurements is achieved.  In light of the deuteron's evident fragility,
information obtained from this analysis may be useful in establishing
freeze-out volumes and help in heralding the presence of high-density
phenomena in a baryon-rich environment.
\end{abstract}   


\section{Introduction}  

Several coalescence models 
\cite{ButlerPearson,Schwarz,Remler,Bond,Mekjan,Kapusta,GFR,Mr,DB,DHSZ,KahanaDover},
have been proposed for
calculation of cluster production in heavy ion collisions. In this paper we
examine the use of such modeling for deuterons only, and with particular
reference to existing Si+Au data at AGS energies. We demonstrate that it is
necessary to understand something of the quantum mechanical aspects of
coalescence in order to extract the absolute magnitude of cluster
yields. Given this, it may then also be possible that information on the size
of the ion-ion interaction region, complementary to that from HBT \cite{hbt},
will flow from a study of deuteron production. It must be emphasized that the
interaction region or ``fireball'' spatial extent can only be gathered from
knowledge of {\it absolute} deuteron yields and is in general lost if, for
example, the acceptances for formation in position and momentum are adjusted
to make theoretical yields agree with experiment \cite{KahanaDover,Kumar et
al} and/or the quantal aspects are ignored as in the ``cutoff''
models described in what follows.  Most interesting would be the case of
disagreement between an improved, self-consistent, cascade calculation and
experiment. One would like to conclude, in the presence of such a
discrepancy, that the fireball lives significantly longer (or shorter) than
the cascade suggests.  Our development can be  usefully
compared to a study by Koonin\cite{Koonin1977} of the nucleon pair
correlation function generated in heavy ion collisions. The deuteron provides
the best cluster for present purposes because, although the simplest, its
spatial dimensions are still quite comparable to those expected for ion-ion
interaction regions. The use of larger clusters may complicate the theory
without adding much to use of coalescence as a probe of unusual
medium effects. We emphasize that the rapidity region considered in this
work, both theoretically and experimentally, avoids the target and
projectile points where confusion with ``boiled'' off clusters might occur.
Perhaps more importantly, the deuteron is weakly bound and
its final materialization most likely occurs only after cessation of strong
interactions for the coalescing nucleon pair. Thus a factorization of the
calculation into a piece arising from the cascade, i.e. the pair nucleon
distributions, and one arising from the quantum coalescence, is very probably
a realistic description.
\clearpage

Since bound state formation is sensitive to the presence of even a slight
correlation between the space-time and momentum vectors of the two coalescing
nucleons, one can also extract from deuterons evidence for collective
motion, i.e. hydrodynamic flow.  The latter analysis may be complicated by
the presence of "preformed" deuterons in both target and projectile regions of
rapidity.  We in fact examine the content of our produced deuterons to
establish whether the two nucleons come from the same or different initial
nuclei. A sizeable correction due to deuteron formation was indeed necessary
for evaluating the level of nearly forward protons generated in Si+ Pb
collisions at 14.6 GeV \cite{ARC814,E814}, wherein a very rudimentary version
of coalescence was used. However, and not surprisingly, there is in the final
analysis a strong, and useful, correlation between deuteron parentage and
rapidity.

As noted, the weak binding of the deuteron can also be used to advantage,
permitting one to factor production into an initial stage in which the event
simulator, in this case ARC \cite{ARC,ARC814}, generates the single nucleon
distributions, and a second stage in which the coalescence takes place. The
separation between these stages is reasonably well defined for the deuteron;
it is marked by the last collision of both of the combining nucleons, i.e. at
``freezeout''. Earlier formation, or at least survival, of the weakly bound
deuteron is unlikely. It is in just such circumstances that the more global
coalescence models can be best expected to work. Just what one means by
``last'' collision is, however, also subjected to some scrutiny here. The
simulation is cut off below some cm  energy for colliding (not coalescing)
particles, and the sensitivity to this cutoff is tested. One might comment at
this point that coalescence of anti-deuterons in ion collisions should be
very similar to that of deuterons. The anti-nucleon and nucleon distributions
might differ appreciably, the anti-particles being in some senses
surface--constrained by annihilation\cite{Mr2}, but the freezeout of
anti-deuterons is again dominated by the low binding energy. We will examine
such exotic clusters in future work, although we include some discussion
here.

Another advantage of weak deuteron binding is that more ``microscopic'' but
harder processes, such as concomitant final state $\pi$ production, are
considerably less likely than the soft coalescence. Given anticipated
limitations on the level of accuracy in both data and our present theory, we
ignore these auxiliary channels.

The coalescence model depends crucially on the space and momentum
distributions of neutrons and protons. To be as precise as possible we use
the relativistic cascade ARC \cite{ARC,ARC814}, which has been very
successful in describing and predicting \cite{Hipags} the measured nucleon
spectra in several AGS experiments. As we will show in Sec.~3, the interface
between the ARC code and the coalescence model is relatively simple, but
still requires some design choices.

Another important ingredient is the quantum mechanical ``device'' used to
marry the ARC distributions to the bound wave function. Ideally this would be
done at the microscopic level, with perhaps interaction with a ``field'' or a
third object placing the deuteron on-shell. We will, for obvious reasons,
stick with the coalescence model.  We will in fact perform three related
calculations to test the quantal and spatial features of the coalescence
modeling:

\begin{itemize} 
\item Static: A calculation in which neutron, proton and
deuteron wave packet sizes are set externally and globally. 
\item Dynamic: A calculation in which the sizes are determined for each
coalescing pair during the cascade.
\item Wigner: A calculation using the Wigner function formalism. We consider 
two variants:
 	\begin{description}
		\item[(1)] Wigner as generally implemented (Standard Wigner).
                \item[(2)] Wigner as introduced below (Quantal Wigner).
 	\end{description}
\end{itemize}

Our standard calculation, referred to hence as ARC Dynamic, is the second of
these because of its physical basis and because it alone yields the
possibility of a parameter free determination of absolute deuteron yields.
The Wigner characterisation should itself be divided into two limiting cases,
discussed in some detail in what follows.  A ``generalized'' Wigner procedure
is precisely equivalent to the first or Static wave packet scheme. Once one
has factorized the calculation into two parts, the overlap integral
estimating deuteron formation may be subjected to the Wigner
transformation. The result is a convolution of the deuteron Wigner transform
with the neutron and proton Wigner functions. An ``exact'' Wigner simulation
would then assign wave packets to the two nucleons, taking account of the
central or average position and momenta of these packets. If for example one
selects some appropriate smearing size for the single particle wave functions
and performs the requisite convolution with the known bound-state deuteron
Wigner function, then the Wigner procedure is identical to the Static
approach, both incorporating the quantum mechanics inherent in the overlap
integral.

However in the "standard" Wigner treatment, generally employed
\cite{Remler,GFR,DB}, one attempts to fix both the (classical) momentum and
position of cascading particles, a procedure at variance with the precepts of
quantum mechanics. This constitutes a definite approximation to the quantal
treatments presented above. One might very well wish to compare the results
of this Standard Wigner, with apparently no quantal smearing specified for
the initial nucleons, with those of the generalised Wigner with some
smearing $\sim 1fm$. We compare these approaches with each other and with
Dynamic coalescence.  This comparison (see Figs.~5,6) exhibits appreciable
disagreement in absolute deuteron yields, with interestingly the largest
divergence between the two Wigner calculations.

Our results are predicated on the factorisation referred to above,
i.e. coalescence should occur only after all reactions have ceased for the
participant np pair.  We require then a knowledge of both the relative
distance of the two nucleons {\it and of} the spatial extent of their wave
packets at the freeze-out time of each pair. The distribution of the former
is assumed given by the cascade; the latter ostensibly follows from the
quantum mechanical history of the individual nucleons somewhat before and
during coalescence. Fortunately, deuteron formation seems to be sensitive
only to the size of the packets and not to finer details.

The internal deuteron wave function is well known\cite{deuteronwf,chargedradii}. We demonstrate below that
only moderate sensitivity exists to the root mean square radius $r_d$ of the
deuteron and hence we do not expect great dependence on the specific form of
wave function. We test this sensitivity by using as a measure of the deuteron
bound state size both the charge radius determined from electron scattering
and the ``point'' radius obtained after removing the finite proton charge
radius.  In Static the wave packet sizes are externally fixed, remaining the
same for an entire nucleus-nucleus collision. For Dynamic the wave packet
sizes are determined separately from the environment of each pair. Causality
suggests that a given nucleon should be affected only by hadrons in its past
light cone. This will be kept in mind when determining the wave packets of a
coalescing pair. The potentially new element that wish to highlight
is the physics represented by the parameter(s) describing the spatial extent
of the nucleon wave packets.  These parameters, largely ignored in earlier
calculations or hidden in the choice of Wigner function, should perhaps be
related to the size of the interaction region generating or ``preparing'' the
single nucleon distributions \cite{Koonin1977}. One might have imagined the
Wigner formalism would be tied to the use of a small wave packet size or
smearing. Figures~3 and 4 would seem to suggest otherwise.

In earlier work \cite{KahanaDover} we considered the  formation of a
weakly bound (1-15 MeV) $\Lambda\Lambda$, our version of the H-dibaryon. 
There, however, a coalescence
prescription was used confining bound state production to baryon pairs
contained within a certain region in the relative six-dimensional phase
space. The radii of this allowed region were related to the relative position
and momentum content of the true (but unknown) H wave function. The
normalisation for the di-hyperon cross-section was obtained, however, by
comparing a similar cutoff calculation for the deuteron to existing data
\cite{E802summary}. It is such a cutoff model that we intend to improve on
here.

Section 2 contains a rudimentary coalescence theory, including a brief
statement of the relation between the wave packet and Wigner
approaches. Section 3 describes implementation of our preferred approach,
ARC Dynamic, within the framework of ARC, and in Section 4 we examine our
results with an eye to understanding the quantal effects introduced above and
to comparing these results to published measurements \cite{E802summary}.  In
general the dynamical approach does surprisingly well in reproducing the
overall experimental Si+Au measurements, both detailed transverse-mass
spectra and significantly, the absolute magnitudes.  The confidence generated
by this result encourages us to pursue the matter further, to the point of
suggesting that unexpected measured deuteron production might signal unusual
behaviour.

Section~5 contains highly instructive details of the space-time and momentum
evolution of coalescing pairs within the Dynamic simulation which can lead to
some estimates of freezeout sizes at the time of coalescence.  Section~6
contains a brief summary. One must keep in mind that to at least some extent
the success of the cascades, in general and in their treatment of cluster
formation, depends on the considerable degree of averaging taking place even
in a single ion-ion event. Nevertheless, the theory produces a credible
picture of the physics, as will become clear in what follows.

\section{Coalescence Theory}

\subsection{Overlap Ansatz}

The ``bare-bones'' coalescence model first proposed historically \cite
{Schwarz}, was one in which only the relative momentum of combining particles
need be within a predetermined range. Strictly speaking, such a limit is only
valid if the particle source is small in comparison to the final bound state.
Effectively, one is then using plane waves for the single particle wave
packets. In its simplest incarnation the coalescence model forms a deuteron
from a proton and a neutron if their relative momenta are within a certain
{\it capture radius} $p_0$, comparable to the deuteron's momentum
content. The deuteron cross section can then be computed
(non-relativistically) in terms of the proton and neutron cross sections
as
\begin{equation}
\frac{d\sigma_d}{dp^3} ({\bbox{p}}) = \frac{3}{4}\frac{4\pi p_0^3}{3}
\frac{d\sigma_p}{dp^3} \big( {\textstyle\frac{{\bbox{p}}}{2} } \big)
\frac{d\sigma_n}{dp^3} \big( {\textstyle\frac{{\bbox{p}}}{2} } \big)
\end{equation}
where $\frac{3}{4}$ is the combinatoric factor for angular
momentum\cite{Mekjan,GFR}.

For pA collisions the omission of any spatial dependence is
perhaps justified, given that the interaction region is small enough for both
deuteron wave function and nucleon density to be taken constant. For ion-ion
collisions the final interaction region might in fact include the entire
smaller nucleus and some account must be taken of spatial dimensions relative
to the deuteron.  Consequently, the momentum capture radius $p_0$, as
extracted from measured proton and deuteron spectra, exhibits an unnatural
variation with target and projectile \cite{Gutbrod,Lemaire}. We are
attempting to elaborate upon this observation here.

In examining the H-dibaryon and deuteron formation in ion
collisions\cite{KahanaDover}, the present authors extended this ``cutoff''
coalescence to include a constraint on both relative spatial and momentum
separations
of combining baryons, i.e. one then defined an allowed six, rather than
three-dimensional phase space region. By normalising to measured deuteron
yields in Si+Au collisions, we hoped to remove  ambiguities in the
H-yield. These cutoff calculations can not determine absolute cluster 
magnitudes, incorporating as they do a strong dependence on the six-dimensional
phase-space volume in which coalescence takes place.  The finite size of the
deuteron may presage an interesting dependence of production on impact
parameter, over and above that due to the single nucleon distributions.
 
The naive-cutoff coalescence prescriptions have major shortcomings. In the
first instance the purely momentum treatment contains no spatial information,
while the extended space-momentum version introduces such information in an
ad hoc fashion. This cannot be easily fixed without violating quantum
mechanics which evidently forbids the simultaneous use of precise momentum
and space coordinates.  Further, since the quantum mechanics of the formation
is absent, so is the actual deuteron wave function, which might perhaps be of
essential importance in the microscopic process.  These shortcomings can be
avoided within a rudimentary quantum mechanical model
\cite{GFR,DB,DHSZ,Koonin1977} which assumes that neutron and proton are
described by wave packets of width $\sigma$ localized in space around
$\bar{\bbox{x}}_i$ and in momentum space around $\bar{\bbox{p}}_i$:

\begin{equation}
\psi_i({\bbox{x}}) = \frac{1}{(\pi\sigma^2)^{3/4}}
	\exp\Big( -\frac{({\bbox{x}}-\bar{\bbox{x}}_i)^2}{2\sigma_i^2} \Big)
	\exp\Big( i\bar{\bbox{p}}_i{\bbox{x}} \Big)
\label{wavepacket}
\end{equation}
 
The two ARC approaches, Dynamic and Static are distinguished at this point by
choice of the size parameters $\sigma_i$ for the neutron and proton. In the
latter this choice is simple, a single global value for all pairs and events.
It is surely a simplification to imagine one spatial parameter describes all
nucleons partaking in an ion-ion collision. However, considering the
complexity of the interactions a reasonable averaging may result. The
determination of $\sigma$ for Dynamic coalescence from the pair history is
described in the next section.

We write the deuteron wave function as a product of its center-of-mass motion
and its internal motion
\begin{equation}
\psi_d( {\bbox{x}}_1, {\bbox{x}}_2 )
	= \Phi_{\scriptstyle \bar{\bbox{P}},\bar{\bbox{R}}}( {\bbox{R}} ) \, \phi_d({\bbox{r}})
\end{equation}
where
\begin{equation}
\Phi_{\scriptstyle \bar{\bbox{P}},\bar{\bbox{R}}}( {\bbox{R}} )
        = \frac{1}{(\pi\Sigma^2)^{3/4}}
	\exp\Big( -\frac{({\bbox{R}}-\bar{\bbox{R}})^2}{2\Sigma^2} \Big)
	\exp\Big( i\bar{\bbox{P}}{\bbox{R}} \Big)
\end{equation}
and
\begin{equation}
\phi_d({\bbox{r}})
         =(\pi \alpha^2)^{-3/4} \exp(-r^2/2\alpha^2)
\end{equation}
In the above ${\bbox{R}} = \frac{1}{2}({\bbox{x}}_1+{\bbox{x}}_2)$ and
${\bbox{r}}={\bbox{x}}_1-{\bbox{x}}_2$, and we have allowed for, but do not
exploit, the possibility that the deuteron center of mass and initial
neutron(proton) wave packets are described by different size parameters. A
natural assumption, which we will make here for simplicity, is that during
coalescence the two body cm motion is unaltered, leading to
$2\Sigma^2=\sigma^2$. In fact it is a rather gentle interaction with some
third particle which puts the deuteron on shell and some small dependence on
the center of mass coordinates should remain. Given that the deuteron is
weakly bound, we assume the latter is small.

We could improve on the choice of relative wave function for the deuteron,
reducing the transparency of the calculation somewhat.  For reasonably
central collisions with massive nuclei the single gaussian should be
adequate, especially in light of the other simplifications being made.  For
the most extreme peripheral collisions, where the fireball size might rival
or be even less than that of the deuteron, this might cause a problem.  One
could test this point with an improved wave function, most easily by fitting
several gaussian terms to say an existing deuteron wave
function\cite{deuteronwf}. The question, here, is whether or not the
coalescence is sensitive to higher moments of the relative motion, and not
just to the rms radius. Probably, in view of the limitations of our modeling
and the wholesale averaging in the ion-ion collision, such fine points are
not significant. We test the sensitivity by varying the deuteron radius
parameter somewhat, and find little effect (see Figs.~11,12).
\clearpage

The coalescence probability, or deuteron content of the two particle wave
function, can now be computed from the squared overlap

\begin{equation}
C({\bbox{x}}_n,{\bbox{k}}_n;{\bbox{x}}_p,{\bbox{k}}_p)
= \vert \langle \psi_n\psi_p \vert\Phi_{\scriptstyle \bar{\bbox{P}},\bar{\bbox{R}}}
          \, \phi_d\rangle\vert^2.
\label{Coverlap1}
\end{equation}
With gaussian packets throughout this yields:
\begin{equation}
\vert \langle \psi_n\psi_p \vert\Phi_{\scriptstyle \bar{\bbox{P}},\bar{\bbox{R}}}
          \, \phi_d\rangle\vert^2
= \left( \frac{4\nu}{\sqrt{2}\mu}\right)^3
\exp\left( -\frac{\nu^2 ({\bbox{k}}_n-{\bbox{k}}_p)^2}{2}\right)
\exp\left( -\frac{({\bbox{x}}_n-{\bbox{x}}_p)^2}{\mu^2} \right),
\label{Coverlap2}
\end{equation}
where $\mu^2=(2\sigma^2+\alpha^2)$, and $\nu=\frac{\alpha\sigma}{\mu}$.

We now assume that the (classical) distribution functions $f_p$ and $f_n$
of protons and neutrons, in so far as they can be obtained from a classical
cascade, actually describe distributions of wave packets centered at the
cascade particle positions and with momenta necessarily spread about the
cascade values.  Again, the choice of size parameters $\sigma_i$, $i=p,n,D$,
for the wave packets is handled differently in our two protocols, uniformly
for all coalescing pairs in one ion-ion event for Static and from the history
of each pair for Dynamic.  Including a factor of $\frac{3}{4}$ for spin the
number of deuterons, and neglecting pairs lost to higher clustering, the
deuteron number is then

\begin{equation}
n_d = \frac{3}{4}
	\int d\bar{\bbox{x}}_n d\bar{\bbox{k}}_n  f_n(\bar{\bbox{x}}_n,\bar{\bbox{k}}_n)
	\int d\bar{\bbox{x}}_p d\bar{\bbox{k}}_p  f_n(\bar{\bbox{x}}_p,\bar{\bbox{k}}_p)
	\; C({\bbox{x}}_n, \bar{\bbox{k}}_n; \bar{\bbox{x}}_p, \bar{\bbox{k}}_p),
\label{ndoverlap}
\end{equation}
which, given the above interpretation of cascade distributions, may be written
\begin{equation}
n_d = \frac{3}{4}
   \sum_{ij} C({\bbox{x}}_{n_i}, {\bbox{k}}_{n_i}; {\bbox{x}}_{p_j}, {\bbox{k}}_{p_j}),
\end{equation}
with the sum extending over all appropriate pairs {ij} in the cascade. The
quantum fluctuations of individual nucleons or of the deuteron center of mass
motion are built in through the wave packets. The situation is different as
we will see for the ``Standard'' application of the Wigner formalism, but not
necessarily in a more straightforward realisation of the latter.

It should be noted, in agreement with References~5 and 7, that there is no
isospin factor of one-half in this expression \cite{Kumar et al}; neutrons
and protons can indeed be treated as distinguishable. If one were to use an
isospin formalism it would be necessary to symmetrize the cascade input to
coalescence with respect to the np pair, and the result inevitably is the
same with the apparent factor of one-half compensated by a symmetry factor of
two.  The symmetrisation must be imposed externally since the classical cascade would
never yield both $n(k_1)p(k_2)$ and $n(k_2)p(k_1)$.

\subsection{Equivalence to Wigner Function Formalism.}

The equivalence of the Wigner\cite{GFR,DB} and overlap coalescence is self
evident under a factorization hypothesis, i.e. if one separates the cascade
generation of single nucleon distributions from deuteron formation. One can
re-express Eq.~(8) for the deuteron yield in terms of Wigner functions
$f_i^W$ for the initial neutron, proton and final deuteron:
\begin{equation}
n_d^W = \frac{3}{4}
	\int d{\bbox{x}}_1 d{\bbox{p}}_1  f_p^W({\bbox{x}}_1,{\bbox{p}}_1)
	\int d{\bbox{x}}_2 d{\bbox{p}}_2  f_n^W({\bbox{x}}_2,{\bbox{p}}_2)
	\; f_{deut}^W({\bbox{x}}_1, {\bbox{p}}_1; {\bbox{x}}_2, {\bbox{p}}_2).
\end{equation}
These functions $f_i^W$ are simply transforms of appropriate density
matrices, in the fashion:
\begin{equation}
f_i^W({\bbox{x}},{\bbox{p}}) = 
   \int d{\bbox{\eta}}  \rho_i({\bbox{x}}-\frac{{\bbox{\eta}}}{2}, {\bbox{x}}+\frac{{\bbox{\eta}}}{2} )
\exp(-i{\bbox{p}}\ . {\bbox{\eta}}). 
\end{equation}
For pure states we may write the density distributions in terms of our
previous wave functions as
\begin{equation}
\rho_i({\bbox{x}}, \bar{\bbox{x}} ) = \Psi_i({\bbox{x}}) \Psi_i^*(\bar{\bbox{x}}).
\end{equation}
Inserting the densities in Eq.~(12) into Eqs.~(10,11) and performing the required 
integrations results in
\begin{equation}
n_d^W=
  \sum_{ij} \vert \langle \psi_n\psi_p \vert\Phi_{\scriptstyle {\bbox{P}},{\bbox{R}}}
          \, \phi_d\rangle\vert^2,
\end{equation}
which in fact implies the identity
\begin{equation}
n_d^W= n_d
\end{equation}
provided only that the wave functions entering the Wigner transforms are the
same as used in Eqs.~(2,3).  This comes as no surprise; the exact Wigner
transformation in Eqs.~(11,12) has been undone by inserting the density
operators defined through wave packets into Eq.~(10).

Standard Wigner \cite{Remler,GFR,DB} takes another path, specifying the
neutron and proton distributions in Eq.~(10) through
\begin{equation}
f_i^W({\bbox{x}},{\bbox{k}})= \delta({\bbox{x}}-{\bbox{x}}_i) \delta ({\bbox{k}}-{\bbox{k}}_i).
\end{equation}
The treatments thus diverge when one uses this usual, but quantally
disallowed, assumption for the Wigner distributions. In fact, in the case of
a gaussian wave function for the bound deuteron, one can continue to use
Eq.~(9) with the Standard Wigner coalescence with
\begin{equation}
 C({\bbox{x}}_{n_i}, {\bbox{k}}_{n_i}; {\bbox{x}}_{p_j}, {\bbox{k}}_{p_j})
= 8
\exp\left( -\alpha^2({\bbox{k}}_{n_i}-{\bbox{k}}_{p_j})^2\right)
\exp\left( -\frac{({\bbox{x}}_{n_i}-{\bbox{x}}_{p_j})^2}{\alpha^2} \right),
\end{equation} 
where $\alpha$ is the deuteron size parameter, related to the ``point'' rms
radius $r_P$ by
\begin{equation}
(r_P)^2= \frac{3}{2}\alpha^2
\end{equation}                                      
There are evidently no free parameters in this result, any smearing of the
nucleon positions and momenta is hidden and perhaps arises only from event
averaging. In fact the algorithm Eq.~(16) for  Standard Wigner followed
from assuming precise values simultaneously for both space and momentum. This
is clearly evidenced \cite{GFR}  in the factor $8$ in Eq.~(16)  in what is
supposed to be the formula for a probability.  One cannot 
simultaneously define both position and momentum for a cascading
particle. If probabilities for coalescence are not in general large for
small spatial separation, and if one averages sufficiently in each ion-ion
collision, then this distinction may not be numerically too
significant. Nevertheless, it is surely safer to employ Eqs.~(6,7) rather
than Eq.~(16). Also at least part of the important physics lies in assigning
wave packet sizes. Figs.~5,6 comparing ARC to the two Wigner calculations
demonstrate that some price in absolute normalisation of the deuteron yields
must be paid. The relative normalisation of Standard Wigner to ARC Dynamic
changes appreciably between central and peripheral collisions. Clearly,
Static coalescence, and its equivalent partner Quantum Wigner,
contain a size parameter. The dynamic modeling permits this parameter to be
internally estimated, yielding a higher degree of predictability.

We might offer as metaphor for microscopic rendering of coalescence an
analogy with either deuteron stripping in finite nuclei and/or inelastic
scattering. The latter gives one a more directly comparable expression for
the probability displayed in Eq.~(6), but the former, stripping analogy,
could give a more concrete model if pursued. Both formalisms applied to the
system of neutron+comoving nucleons, after the short range interactions
between cascading particles have ceased, suggest using neutron and proton
wave packets defined in the long range field generated by the
comovers. Although it is far from trivial to evaluate this field, our dynamic
approach may be viewed as making a first estimate of its spatial extent. The
numerical results are encouraging. It would seem perhaps only the spatial
extent of the particle wave functions play a role, the details of the field
not being overly significant. We reiterate that this field is weak, long
ranged and would little affect the single nucleon distributions.

\section{Dynamic Coalescence: Implementation into ARC.}

Since the functions $f_p$, $f_n$ in Eq.~(8) describe the distribution of
centers and average momenta of nucleon wave packets as generated by the
cascade, we were led to write for the minimal quantal treatments, i.e. for
ARC Dynamic, Static or for the quantal Wigner
\begin{equation}
n_d = \frac{3}{4}
   \sum_{ij} C({\bbox{x}}_{n_i}, {\bbox{k}}_{n_i}; {\bbox{x}}_{p_j}, {\bbox{k}}_{p_j}),
\end{equation}
where the sum in Eq.~(9) can be restricted to np pairs with fixed kinematics,
e.g.  given rapidity and transverse mass. Indeed, as we have shown above,
Quantum Wigner is just ARC Static with a fixed, likely small, size parameter,
$r_{wp}=1fm$ or equivalently $\sigma=0.817fm$. 

Specifically, our procedure within the simulation is to select pairs of
nucleons, one neutron and one proton, follow their trajectories until both
have ceased interacting with other hadrons and then evaluate, within a
Monte-Carlo framework, the possibility of coalescence. Should this occur, the
nucleons are removed from the particle lists and replaced by the appropriate
deuteron. Although in some low probability coalescence events one may find
appreciable non-conservation of energy, conservation of momentum is
guaranteed. The effect of this on the results is necessarily small and the
non-conservation of energy limited to a few hundred MeV in a Au+Au collision.
This is repeated for all choices of the pair within one ion-ion collision. As
stated previously, it is assumed that deuterons formed before the cessation
of interactions will not survive. Not only does the simulation, event by
event, generate the nucleon precursor average positions and momenta, it also
can guide us towards an evaluation of the size of their wave packets. The
interaction history of the nucleons before freezeout can be used to estimate
a radius for the fireball, and hence yield a value for the parameters
$\sigma_i$, or since we have taken the neutron and proton size parameters
equal and related these to the deuteron center of mass size, for a single
$\sigma$. This $\sigma$ will vary with the environment of the selected pair,
impact parameter and perhaps also with the kinematics of the reaction. This
better understanding of the relativistic and hence spatial aspects of cluster
formation requires a more integrated version of coalescence within the
cascade dynamics.  In the next section we present model calculations and
compare them to existing data.

From Eq.~(7) it is clear that within the assumptions we have made, the
relative position
\begin{equation}
{\bbox{x}}={\bbox{x}}_n-{\bbox{x}}_p 
\end{equation}
and momentum
\begin{equation}
{\bbox{k}}=\frac{{\bbox{k}}_n-{\bbox{k}}_p}{2} 
\end{equation}
are the only classical variables entering the overlap calculation. A central
question then is the choice of neutron and proton wave functions within the
ion-ion simulation. The n,p wave packet product represents an initial two
nucleon wave function, which we imagine prepared at a coalescence time
contemporaneous with the last interaction of both nucleons, i.e. at
$t_c=max(t_n,t_p)$.  The relative position and momentum are then evaluated in
the two particle cm frame, and from these the chance of coalescence
determined.  In the Static case the probability is calculated in Eqs.~(7,8)
with a wave packet size fixed for all pairs. In Dynamic the wave packet size
is estimated separately for each pair, using the distribution of previous
interactants.

It is reasonably evident that only particles in the backward light-cone of
the coalescing pair should define the nucleon and deuteron cm wave
packet size. We draw the light cone at the coalescence point (see Fig.~19),
${\it x}_c^{\mu}=({\it x}_c,{\it t_c})$, whose spatial coordinates lie at the
midpoint of the coalescing np pair in their mutual cm frame.  Moreover,
since coalescence occurs only after freezeout, one must make this
determination as late as possible. The option which suggest itself as most
consistent with these constraints is propagation of the co-interactants as
closely as possible to the light-cone of the coalescing pair (again see
Fig.~19). Alternatively, one could calculate an average position for the
interacting particles in the backward light-cone, or use their initial
positions. We will discuss the numerical effects of alternatives. Spectators
are generally neglected in the calculation of wave packet sizes as are
particles causally disconnected from the coalescence. However, for deuterons
coalescing purely from spectators, the Dynamic calculation of size includes
only spectators. The assumption made here is that the initial nucleus size
determines the wave packet of these essentially undisturbed nucleons. The
wave packet size is equated, in Dynamic, to the rms radius of the interacting
particle region. All quantities necessary for calculation of coalescence
are now available, and there remains only the decision by Monte
Carlo whether or not coalesence actually takes place. There is no double
counting, nucleons forming deuterons are removed from the particle lists.

Static coalescence, for which wave packet sizes are assigned externally,
might simulate Dynamic if for example sizes were adjusted to account for
expected changes in interaction region size, as for example seen in central
and peripheral collisions. We will see that for the systems considered here
i.e. for Si+Au, the different routes followed lead to quantitatively altered
outcomes, at least in overall normalisation. The dynamic simulations seem,
however, to give a consistently accurate picture of existing AGS experiments.
 
\section{Results}

Comparison is made between experiment and coalescence theory for various
choices, Static, Dynamic or Wigner. One might expect the Standard Wigner to
be essentially equivalent to the static theory for a small wave packet radius
assignment, perhaps $\sim 1$ fermi. This, as we shall see, is not so.  An
overall picture of the differences that arise is exhibited in Figs.~5,6 for
Si+Au.

The ARC Dynamic $\frac{dN}{dy}$ spectra in Fig.~4 constitute a comprehensive
normalisation of deuteron production.  Perhaps because the cascade seems to
work so well for absolute deuteron yields as well as for the detailed
transverse-mass and rapidity spectra, one can eventually extract information
about the interaction region. Later, in Figs.~13 to 17, we tie the parentage
of the deuterons to their emerging rapidity.  For existing E802 data on Si+Au
the less disrupted target nucleons play a large role, especially at small
laboratory rapidity. This will not be the case for a central Au+Au
collision where dominantly, fully interacting neutrons and protons coalesce.
We await more comprehensive data for the gold projectiles to illuminate what
might be the most interesting aspects of this study.

We begin with deuteron production for the reaction Si+Au at 14.6
GeV/c. Figures~1 and 2 contain a principal result of this paper, a comparison
between ARC Dynamic  and AGS E802 data \cite{E802summary}, for
transverse-mass proton and deuteron spectra obtained both peripherally and
centrally. Figures~3 and 4, containing rapidity spectra, are essentially
obtained from Figs.~1 and 2 by integration, although in the case of the
experimental spectra some care must be taken to define this integration. The
experimental triggers defining centrality and peripherality
\cite{E802summary} are imposed in the theoretical analysis. There is less
dependence on these triggers for central than for peripheral simulation. In
the latter case E802 has used a different, lower, ZCAL\cite{E802summary} cut
for the most forward angle slice. We have employed a single average cut and
attempted to compensate for the forward trigger in that fashion.

The wave packet size parameters used for  Dynamic are determined within
the simulation, separately for each pair. The deuteron internal, or relative,
wave function is defined by:
\begin{equation}
 \alpha= 1.76fm
\end{equation}
fixed to yield the correct electron scattering
radius\cite{deuteronwf,chargedradii}
\begin{equation}
     r_d= (3/2)^{1/2}1.76fm=2.15fm.
\end{equation}
If one corrects for a finite proton charge radius, $r_p\sim 0.8fm.$, the
appropriate point nucleon distribution is described by $r_d=1.91$fm and
$\alpha=1.56fm$.  We compare yields with both choices of deuteron radius, to
test for sensitivity to the internal deuteron wave function.

Clearly, on a global level, dynamic coalescence does very well indeed. It is
difficult to isolate any systemic discrepancy between measurement and
simulation although differences, generally $\sim10\%$, are on rare occasions
as large as 30-50\%.  Just to what degree these very reasonable theoretical
descriptions of the experimental data, are subject to assumptions and choice
of "parameters" we explore below. We emphasize that the calculated
deuteron and proton spectra are absolutely normalized by the
cascade dynamics with no free parameters; by "parameter" we here mean 
quantities  like $r_d$. These results then suggest the present approach to
coalescence may add a useful tool in the search for interesting medium
dependences. Given the quadratic dependence of deuterons on individual
nucleon distributions, one must of course do reasonably well quantitatively,
in order to ascribe apparent deviations from experiment to interesting medium
dependence. Surprisingly, theory-experiment differences are smaller for
central collisions for which better quality data exists but in which physics
not described in the cascade is more likely present.

No matter how one extracts inverse slopes, or ``temperatures'', from the
experimental or theoretical data, Fig.~1 and Fig.~2 attest to only small
differences between measurement and calculation. For Si+Au, ARC in the dynamic mode closely
reproduces the dependence on transverse mass seen experimentally, in both
shape and magnitude. Figs.~3 and 4 highlight the accord between theoretical
and measured overall magnitudes. 

The E802 rapidity distributions are obtained by fitting a
single exponential to the $m_t$ spectra and after extrapolation into
unmeasured regions. The theoretical $\frac{dN}{dy}$'s are directly integrated
without any such fitting or extrapolation. To the extent that the transverse
spectra are single exponentials, only small additional differences are
introduced in the rapidity spectra. Nevertheless, much larger apparent
differences may be present for the ``temperatures'' extracted from the
experimental data. In a few instances, for the highest rapidities cited by
the E802 collaboration, the deuteron data quality does not support a reliable slope
determination. It is probably better to just compare simulation and
measurement directly and in the case of good agreement to use the theory to
extract a ``temperature''.  In any case the simulations present temperatures
close to experiment in both magnitude and in kinematic
dependences.  The deuteron slopes depart somewhat from the limit of gaussian
convolution of neutron + proton $m_t$ distributions, giving on occasion
higher than naively expected temperatures, but this feature is well mapped in
the cascade.

One should note again that our comparison with experiment begins at a
laboratory rapidity of 0.5 and stops well short of the projectile
rapidity. In extreme peripheral collisions where the nuclei are only mildly
excited and nucleons and clusters simply boil off there is some question
about pure coalescence. Just at $y_{lab}<0.5$, Figs.~3 and 4 show a
slight hint of calculation 
falling below measurements.
However, examination of Fig.~15, which makes explicit the parentage of
coalesced particles, indicates that by $y_{lab}=0.5$ calculated deuterons are
already dominated by s-i pairs and not by purely s-s. This hint may then be
illusory (see also the caption for Fig.~3). In any case measurements at 
lower rapidity would not be amiss.

Comparison of Wigner type calculations with ARC is seen in Figures~5,6.
There are, as we noted, two approaches to the Wigner formalism. Some quantal
aspects can be retained in evaluating the coalescence overlap, whence Wigner
is identical to ARC Static. In such a case it would seem reasonable to assume
the smearing in the neutron and proton wave packets is small, perhaps near to
1 fermi. This we have labeled ``Quantum'' Wigner in Figs.~5 and 6. The result
is a significant reduction, by close to $50\%$, in deuteron yield. The second
approach is the Standard Wigner which inserts sharp definitions of both
nucleon position and momenta in the calculation of deuteron content. The
result of this ad hoc assumption, at best an approximation to the quantum
dynamics, cannot be compensated for by the later averaging over nucleon
distributions inherent in a single cascade collision, nor by event
averaging. What is evident from these figures is the changing ratio of
Standard Wigner to ARC Dynamic as one proceeds from central to peripheral and
in the latter case especially as one moves towards mid rapidity. As we
indicate in the next section on space-time structure, there are several
components in the coalescing deuterons, spectator-spectator,
interacting-spectator and interacting-interacting. Because of our rapidity
cuts $y_{lab}>0.5$ the s-s plays little role here, but the coalescence
of target based nucleons is still important for the Si+Au system \cite{ARC}, 
at least for measurements at less than mid-rapidity.

In Figs.~7,8,9 and 10 the evolution of Static coalescence with a globally
specified wave packet size is explored. There is no way to assign a unique
radius to the packets, but considering the close relation to the oft used
Wigner paradigm, it is very interesting to pursue this evolution.  Clearly
for both central and peripheral analyses there is an ``optimum'' size. This
is already apparent in the rapidity distributions displayed parametrically in
Figs.~7 and 8, but more evident in Figs.~9 and 10 where the magnitudes for
selected rapidities are plotted against $\sigma$, the parameter entering the
gaussian wave functions.  The position of maximum yield, $\sigma_{max}$,
changes with impact parameter $\sigma_{max}\sim 2fm$  for a central collision
and somewhat smaller, $\sigma_{max} \sim 1.25fm$, for peripheral. These correspond to wave packet
radii of $2.5fm$ and $1.5fm$ respectively. This variation with size is not
insignificant since much of the physics is contained in the absolute
magnitudes. Static calculations made near these sizes, producing the
magnitudes comparable to  Dynamic, also exhibit transverse mass
distributions close to those in Dynamic. Standard Wigner central $m_t$
spectra are again essentially indistinguishable from ARC. However, peripheral
Standard Wigner, for which one might have expected the basic assumptions to
be more valid, produces somewhat higher slopes, i.e. lower temperatures than
seen in ARC or in experiment.

Finally, in contrasting Static and Dynamic we note that use of a static size
parameter choice somewhat below $\sigma_{max}$ (see Fig.~14) yields agreement
between these calculations for the central, but a bit above this value for
peripheral.

We have not presented comparative plots, within Dynamic, for
alternative definitions of the ``past history'' of a coalescing pair, though
this choice in principle determines the important wave size. This is because
the results are essentially identical, at least within the accuracy justified
by present experiments, for a wide variety of alternatives.

Explicit comparison of the effect on deuteron yield of changing the deuteron
internal radius from the point to the charge value is presented in
Figs.~11,12. Similar small variation is found with the standard Wigner form
factor, indicating that use of more sophisticated wave deuteron wave
functions\cite{Kumar et al} is unlikely to significantly change results
within that approach.  Also tested was the sensitivity of theory to the
actual experimental triggers: peripheral spectra are sensitive to changes in
these ``cuts'', central considerably less so.

Finally, we have examined the dependence of our results on the energy cutoff
used to halt the cascade. All present calculations were done using a kinetic
energy lower limit $T_{cut}(cm)=30~ MeV$ for elastic collisions. We reran some
cases, both Wigner and Dynamic, for $T_{cut}(cm)=15$~Mev, finding less than
$10\%$ reductions for the lowest rapidities in central collisions and less at
higher rapidities, while for peripheral collisions slight increases. Within
the accuracy of the present theory these are negligible changes. To do better
would require a much more detailed dynamics, including deuteron breakup and
reformation between the times corresponding to the energy cutoffs..

The distributions of precursor nucleon wave packet sizes extracted in the
dynamic treatment are displayed and discussed in the next section. The
figures in that section also exhibit the parentage of the coalescing pair,
divided into three self-descriptive classes, spectator-spectator,
spectator-interacting and interacting-interacting.

\section{Space, Time and Momentum Structure of Coalescing Pairs}
 
It is of great interest to display the space-time history of pairs near
the freezeout time ${\it t_c}$, for nucleons
which both succeed and fail to coalesce. Figs.~13-17 contain such information
for the Dynamic simulations. Fig.~14 succinctly summarizes the information
most immediately relevant to the comparison with E802 data pursued
extensively in this paper. The deuteron data \cite{E802summary} extends over the
rapidity range $0.5\le y_{lab} \le 1.5$ for centrally defined collisions, and
a somewhat more abbreviated range for peripheral. Consequently, in the upper
two graphs of Fig.~14 we
impose a cut $y_{lab} > 0.5$ and display the scatter plot of  wave packet
size $r_{wp}$ vs  rapidity for coalesced np pairs
embedded in a background of  all pairs, for both central and peripheral. The 
lower two graphs in this figure are simple histograms for
these two sets vs $r_{wp}$. One
concludes from the lower graphs that the average $r_{wp}$ for all pairs is
considerably larger than for the coalesced pairs; for either central or
peripheral the overall averages are close to $5fm$, while the coalesced
averages are:
\begin{equation}
\langle r_{wp}(central)\rangle=2.59fm
\end{equation}
and
\begin{equation}
\langle r_{wp}(peripheral)\rangle=1.67fm
\end{equation}
respectively. These values are in good accord with the variation of deuteron
yields vs wave packet size for Static , (see Figs.~9,10) insofar as
such global choices for $r_{wp}$ would result in Static $\frac{dN}{dy}$'s
close to those from Dynamic. Fig.~13, containing similar plots but with no
rapidity cut, tells a different story. The deuterons which fall near
$y_{lab}=0$ arise from target particles experiencing only gentle interactions
and consequently from larger $r_{wp}$. This is especially clear in the
peripheral histograms in this figure, where two distinct groups of pairs are
seen, one at the target ``size'' and another for more strongly interacting
progenitors at a reduced size.

The average $r_{wp}=1.67fm$ for peripheral collisions of Si with
Au is perhaps not much larger than one might have assigned {\it ab initio} in the
Quantum Wigner \cite{privateG}. But the rather steep dependence of yields on the size
parameter in Figs.~9,10 is fair warning that such a choice is better made
dynamically. The central $r_{wp}=2.59fm$ is appreciably larger but still
implies a rather restricted spread in the neutron and proton wave
packets. The dynamic picture is remarkably consistent.

Further interesting information may be gleaned from Fig.~15 on time evolution
of coalescence and from Figs.~16,17 concerning the relative separations of
the coalescing pairs in position and momentum, $\Delta X_{np}$ and $\Delta
P_{np}$.  In particular the range of permissible relative momenta is severely
restrictive, relative momenta larger than 100 MeV/c are rarely seen for a
coalesced pair; a sign of course that the deuteron is a quite low energy
object, weakly held together.

In most figures in this section we have tried to indicate the parentage of
the coalescing pairs, consisting of three groups corresponding roughly to
spectator-spectator, spectator-interacting and interacting-interacting. There
is a further division into target and projectile very closely identified
by rapidity, i.e. from target to mid-rapidity most deuterons consist of two
target particles while beyond mid-rapidity projectile particles dominate. 
There are a few deuterons formed from one target and one
projectile particle at mid-rapidities, but not many.

\section{Summary and Conclusions}

A rather comprehensive investigation of the coalescence model of deuteron
production in a cascade environment has been carried out for a heavy-ion pair
for which extensive data exists, i.e. Si+Au. The principal physical
assumption made is that deuterons survive to maturity only if their component
nucleons have ceased interacting before coalescence. This assumption allows
one to factor the theoretical calculation into a piece depending on the
cascade and a piece depending on the dynamics of coalescence. If, after
factorisation, one is to include quantum mechanics within the formation
dynamics then some knowledge of the spatial and momentum spreading in the
nucleon pair wave functions is required. Dynamic coalescence
provides this knowledge and seems to give a good broad-based description of 
the
measurements. Use of this mechanism to predict Au+Au deuteron yields, at
present AGS energies as in Fig.~18, and for lower energies where this massive
system is more likely to be dominantly equilibrated, then seems
justified. The static paradigm provides a reasonable, and computationally
swifter, description of the data once the global choice for the wave packet
parametrisation is made, but as we have said much of the interesting physics
may lie in this choice. Further, the size parameter $r_{wp}$
is a function of the collision environment, significantly larger for a
central collisions, and certainly varying with rapidity.

Alternative approaches, both Static and Standard Wigner, either requiring or
not the specification of a size parameter, can also apparently give an
acceptable description of the $m_t$ ``angular'' distributions but do not
provide a
completely unified picture of their normalisation. In particular the
evolution of $\frac{dN}{dy}$ from target to mid-rapidities, and again from
peripheral to central is not always correctly tracked. Moreover, the most
interesting deviations from the cascade dynamics, occasioned by high
densities achieved for lengthy times during collision, may well be expected
to appear in overall normalisation. For example plasma formation might 
increase the time till freezeout and consequently the feezeout volume, thus
supressing deuteron formation. 
Excitation functions of deuteron  production
and other interesting observables, will be available in the near future at
AGS energies of 2-8 GeV/c, i.e. just where high densities in a truly
equilibrated system might more reasonably be expected \cite{Hipags}.  One
will then have deuteron data below and above the interesting region and
a drop in yield relative to that expected from  the pure hadronic simulation
would be very interesting indeed. ARC should provide a good predictive
background against which to measure these functions in a search for
unexpected and interesting deviation, i.e. genuine medium effects. We will in
future work present a theoretical analysis of the Au+Au excitation functions
in this energy range.

Anti-deuteron formation can be described by the same picture. A simple rule,
which ought to work well for rapidity distributions, would be to extract the
ratio
\begin{equation}
\rho=
\frac
{\left[\frac{dN}{dy}\right]_d}
{\left[\frac{dN}{dy}\right]_p \left[\frac{dN}{dy}\right]_n}
\end{equation}
from the present calculations and then to construct
\begin{equation}
\left[\frac{dN}{dy}\right]_{\bar d} =
\rho
\left[\frac{dN}{dy}\right]_{\bar p}
\left[\frac{dN}{dy}\right]_{\bar n} ~.
\end{equation}
This evaluation would probably provide adequate numerical accuracy and would
save the considerable computing time required to give passable statistics.
The distributions of anti-protons and anti-neutrons, predicted by the
cascade~\cite{ARCanti} in Eq.~(26) would of course significantly modify the
predictions for $\bar d$'s in both shape and magnitude. It is unlikely,
however, that more information will obtain from such measurements on the
Si+Au system, as exotic as they might be. For Au+Au the increased baryon
densities expected and the tendency of  anti-particles to annihilate might
produce an interesting interplay. ARC calculation of anti-particle production
\cite{ARCanti,E802anti} finds that classical screening of annihilation at low
energies diminishes the density effects, leading to anti-particle rapidity
spectra somewhat narrowed at mid-rapidity, but not drastically different in
shape than other massive produced particles.  Again, new physics might arise from
careful measurement of absolute yields. Both prediction and measurement of
anti-deuteron crossections are complicated by the limits in knowledge of
$\bar p$ production in pp collisions at AGS energies and by the paucity of
anti-deuterons likely to be seen in the data.

In an earlier work \cite{KahanaDover} we considered more massive
clusters and  found it necessary to assign phase-space windows peculiar to each
bound system. There is no barrier to extending the cluster baryon number in
Dynamic, aside from the limitations of computing time. Simulation used for the
design of heavy-ion detectors, for example a possible forward detector at RHIC,
might need to study such massive clusters. We intend to pursue this extension
within a time-saving algorithm. 

This manuscript has been authored under DOE supported research Contract
Nos. DE-FG02-93ER40768, DE--AC02--76CH00016, and DE-FG02-92~ER40699. One of us
(Y.~P.) would also like to acknowledge support from the Alfred P. Sloan 
Foundation.

\clearpage
\begin{figure}
\vbox{\hbox to\hsize{\hfil
 \epsfxsize=6.1truein\epsffile[24 85 577 736]{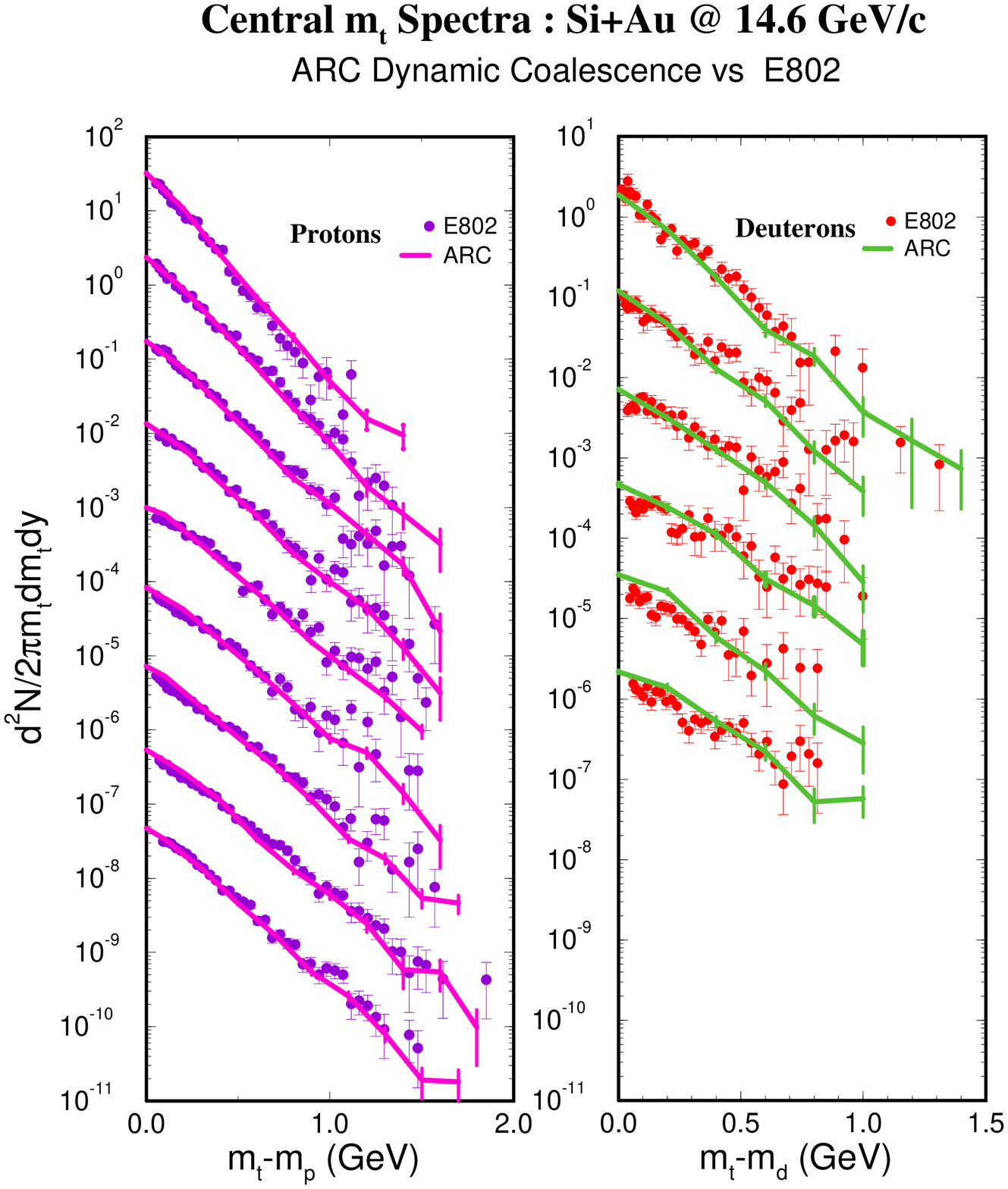}
 \hfil}}
\caption[]{Central Transverse Mass Spectra: ARC simulations are compared to
E802 experiments. Dynamical coalescence determines the wave packet size for
the coalescing nucleon pair, in this case after propagating their interacting
comovers up to the pair light cone. There are then no free parameters in the
theory, the deuteron relative wave function being characterised by the
experimentally determined point size.  There is little variation in these
results with the deuteron size, at least, near the value $1.91fm$ used
here. Using a different prescription for the propagation point, for example
some ``average'' time in the past, also has very little effect. Centrality is
fixed using the E802 specified TMA cut. Little sensitivity to this cut is
evident here. We note the proton spectra in this figure and hereafter are
automatically corrected for deuteron formation, i.e. coalescing protons (and
neutrons) are removed from the cascade. Since the proton spectra enter
essentially quadratically in deuteron formation, the theory is to be judged
also by the matching to singles, a remark which applies to all further
results.}
\label{Fig:one}
\end{figure}
\clearpage

\begin{figure}
\vbox{\hbox to\hsize{\hfil
 \epsfxsize=6.4truein\epsffile[24 85 577 736]{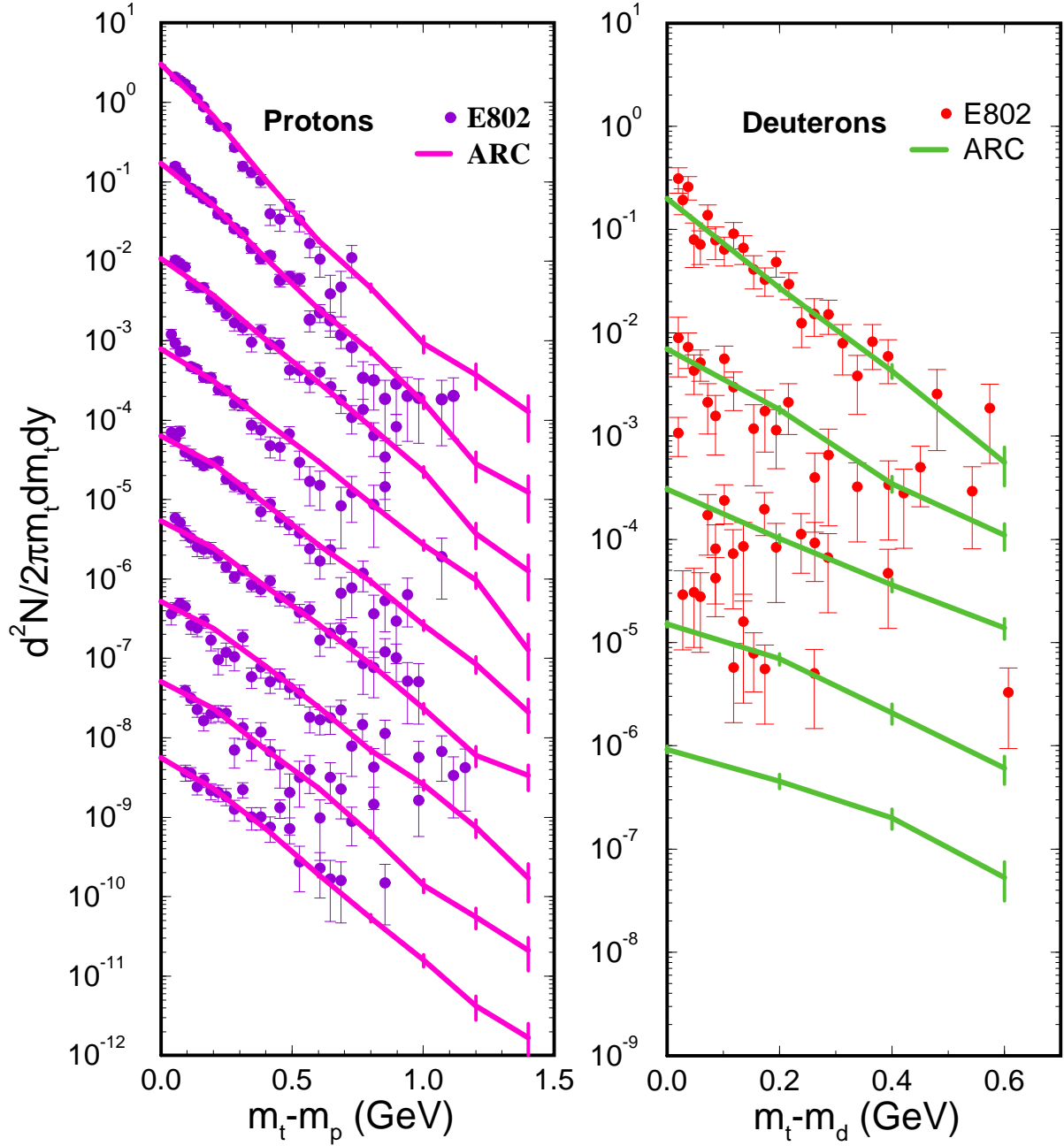}
 \hfil}}
\caption[]{Peripheral Transverse Mass spectra from ARC
dynamical coalescence under the same circumstances as in Fig.~1. Peripherality
is defined using an E802 prescription; there is greater sensitivity to this trigger
than for central collisions. The proton spectra give some indication of the
accord between the theoretical and experimental definitions of the trigger.}
    	\label{Fig:two}
\end{figure}
\clearpage

\begin{figure}
\vbox{\hbox to\hsize{\hfil
 \epsfxsize=6.4truein\epsffile[24 85 577 736]{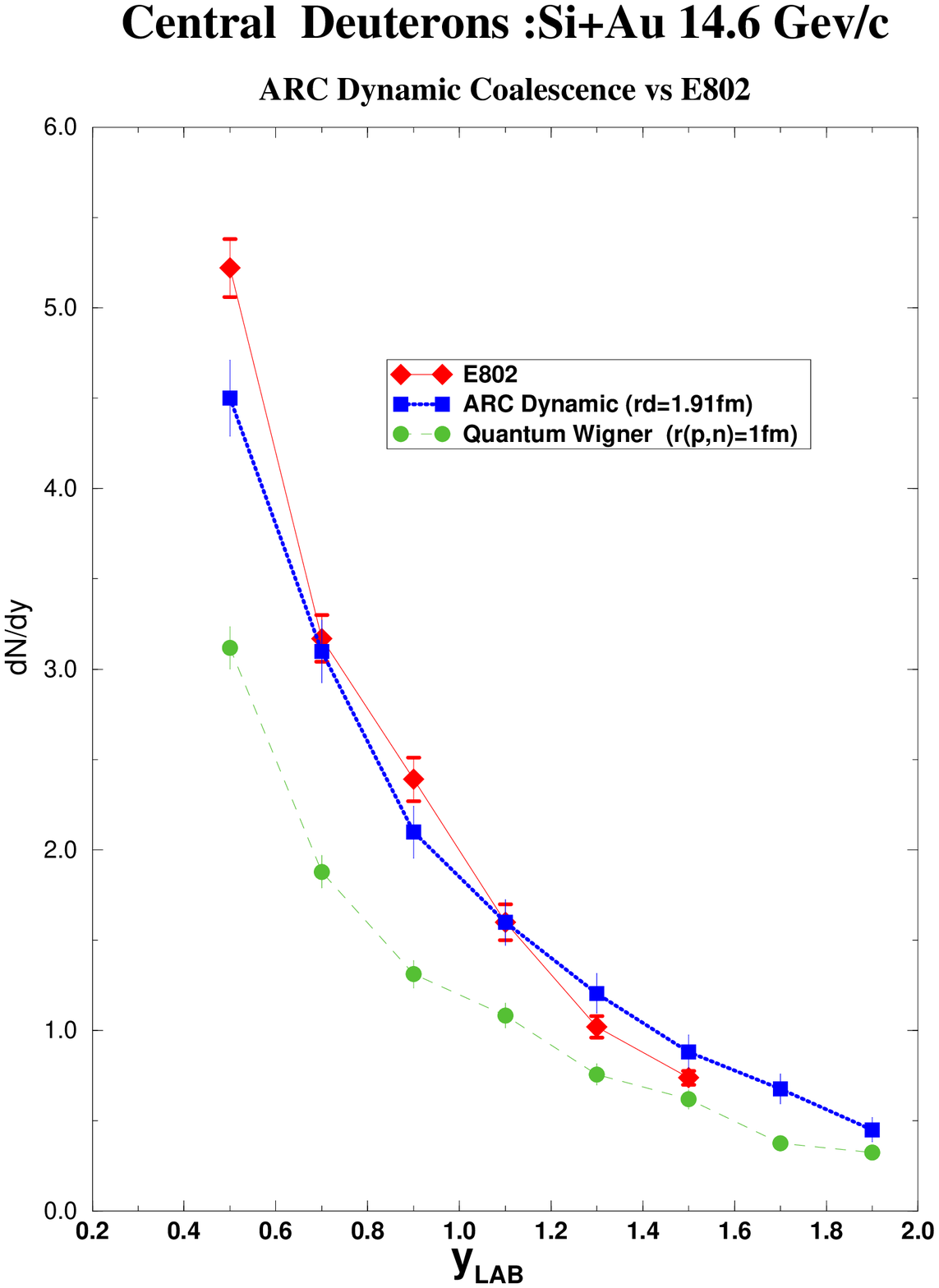}
 \hfil}}
\caption[]{Central Rapidity Distributions from dynamical
coalescence are compared to experimental E802 values. The same, standard,
light-cone prescription as described above and used for the $m_t$ spectra in
Figs.~1,2 has been applied.  The theoretical $\frac{dN}{dY}$ is simply the
integral over the $m_t$ distribution; the E802 value is obtained after
fitting to a single exponential and extrapolation. Differences in the
comparison then arise for deviations  from a simple exponential in Figs.~1,2. For example direct integration of the experimental $m_t$ spectrum
yields a lower value of $\frac{dN}{dY}$ than quoted by E802, resulting in
central value for $y_{LAB}=0.5$ of 4.83 rather than 5.23 and thus bringing 
theory closer to experiment. As indicated in the text the prescription
Quantum Wigner in Fig.~3 is equivalent to ARC Static for a 1 fermi
smearing of the neutron and proton wave packets, i.e. $\sigma=0.817fm$.}
\label{Fig:three}
\end{figure}
\clearpage

\begin{figure}
\vbox{\hbox to\hsize{\hfil
 \epsfxsize=6.4truein\epsffile[24 85 577 736]{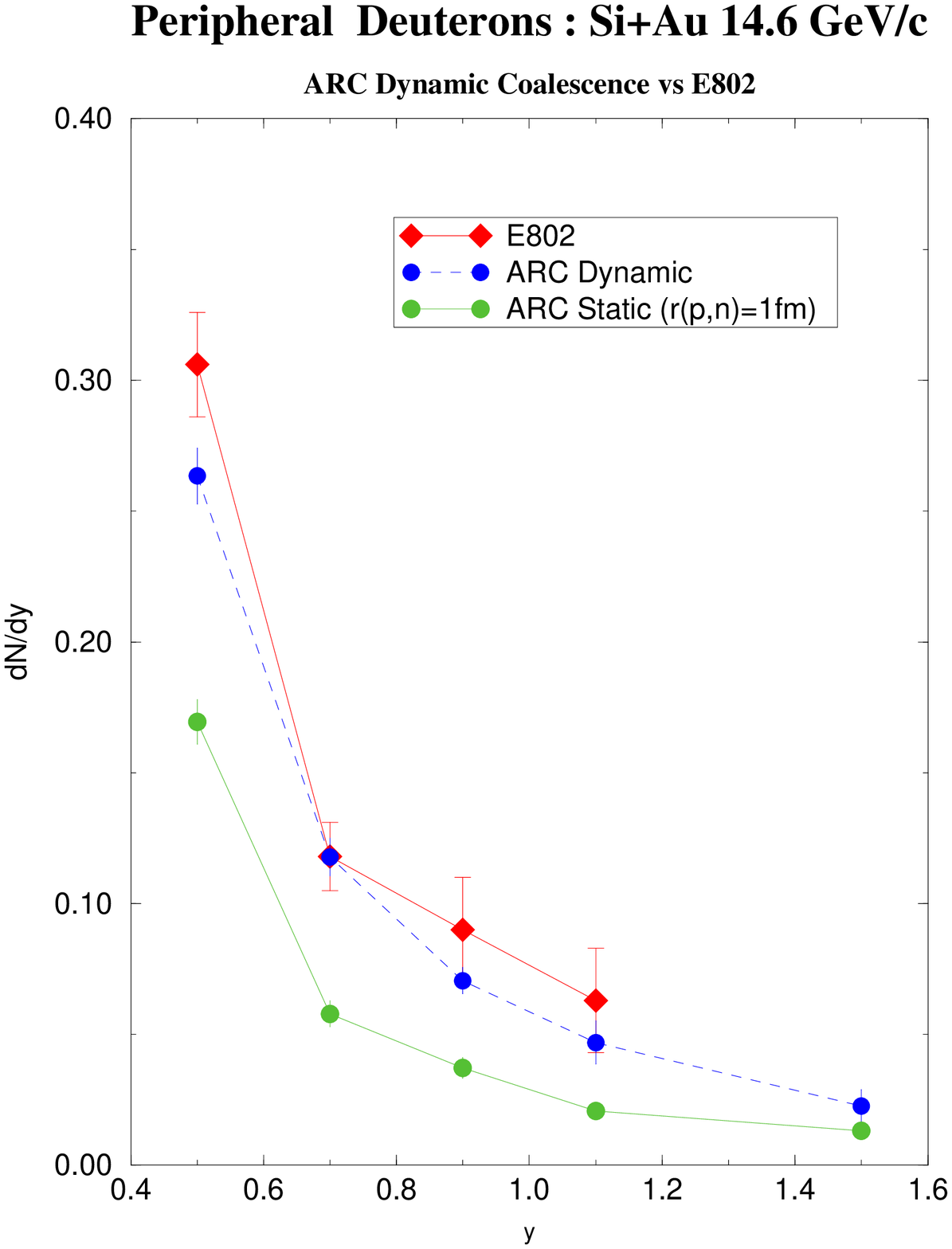}
 \hfil}}
\caption[]{Same as Fig.~3 but for peripheral Si+Au. We note
again here the greater sensitivity to the application of the E802 defined
peripherality using in this case ZCAL.} 
\label{Fig:four}
\end{figure}
\clearpage

\begin{figure}
\vbox{\hbox to\hsize{\hfil
 \epsfxsize=6.4truein\epsffile[24 85 577 736]{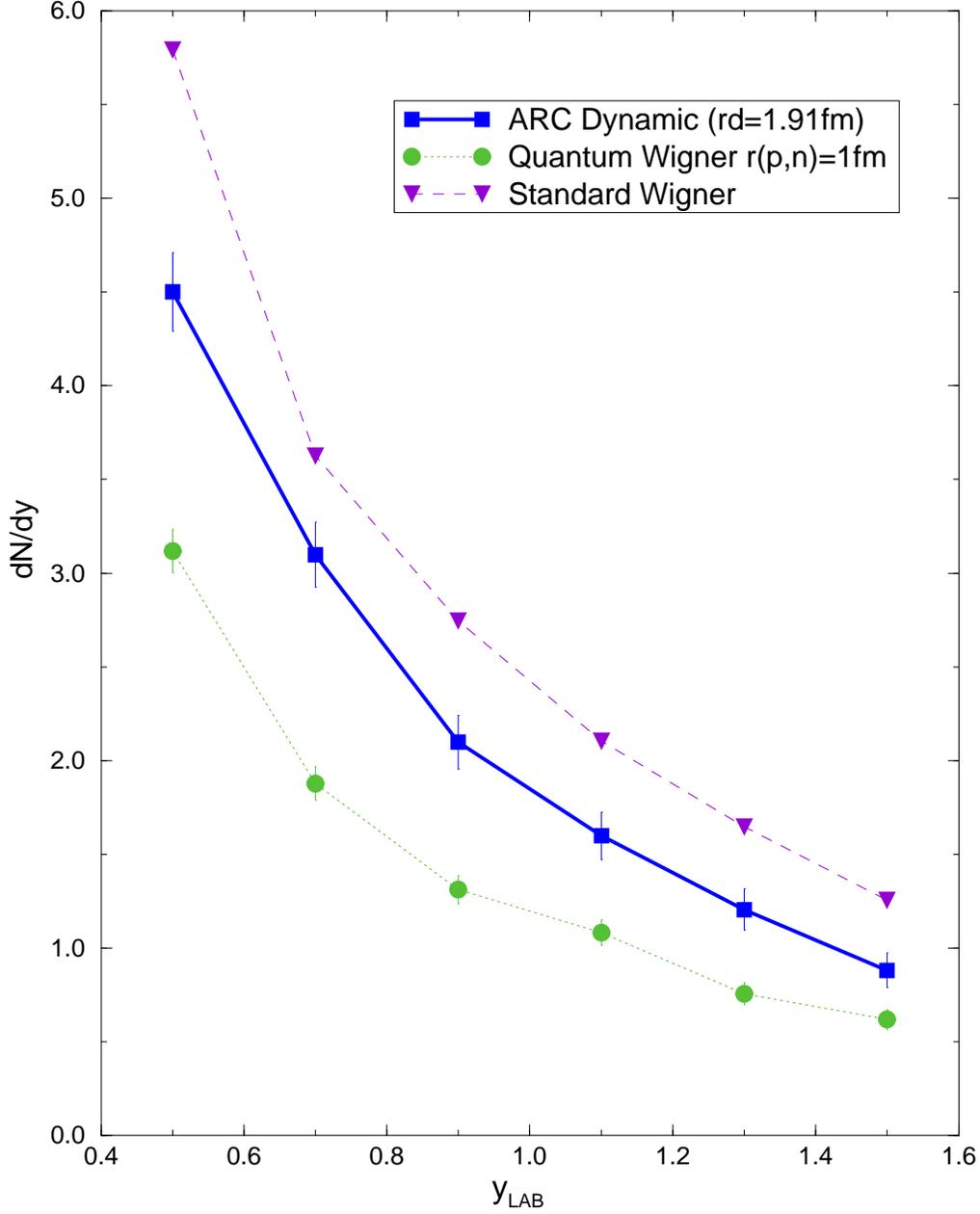}
 \hfil}}
\caption[]{Comparison between ARC Dynamic, Quantum Wigner and
Standard Wigner (obtained using the form factor in Eq.~(16)) for the Si+Au
central rapidity deuterons. Quantum Wigner is identical to ARC Static 
for $\sigma\sim 1.25fm$, i.e. $\langle r_{wp}\rangle\sim 1fm$. Standard Wigner
assumes the classical simulation can be characterised by point
nucleons with sharp momenta. There is a factor of more than two
between these two Wigner coalescence calculations. One notes
also  a similar comparison in Fig.~6 where  Standard Wigner drops, at
mid-rapidity, appreciably below the ARC Dynamic results.}
\label{Fig:five}
\end{figure}
\clearpage

\begin{figure}
\vbox{\hbox to\hsize{\hfil
 \epsfxsize=6.4truein\epsffile[24 85 577 736]{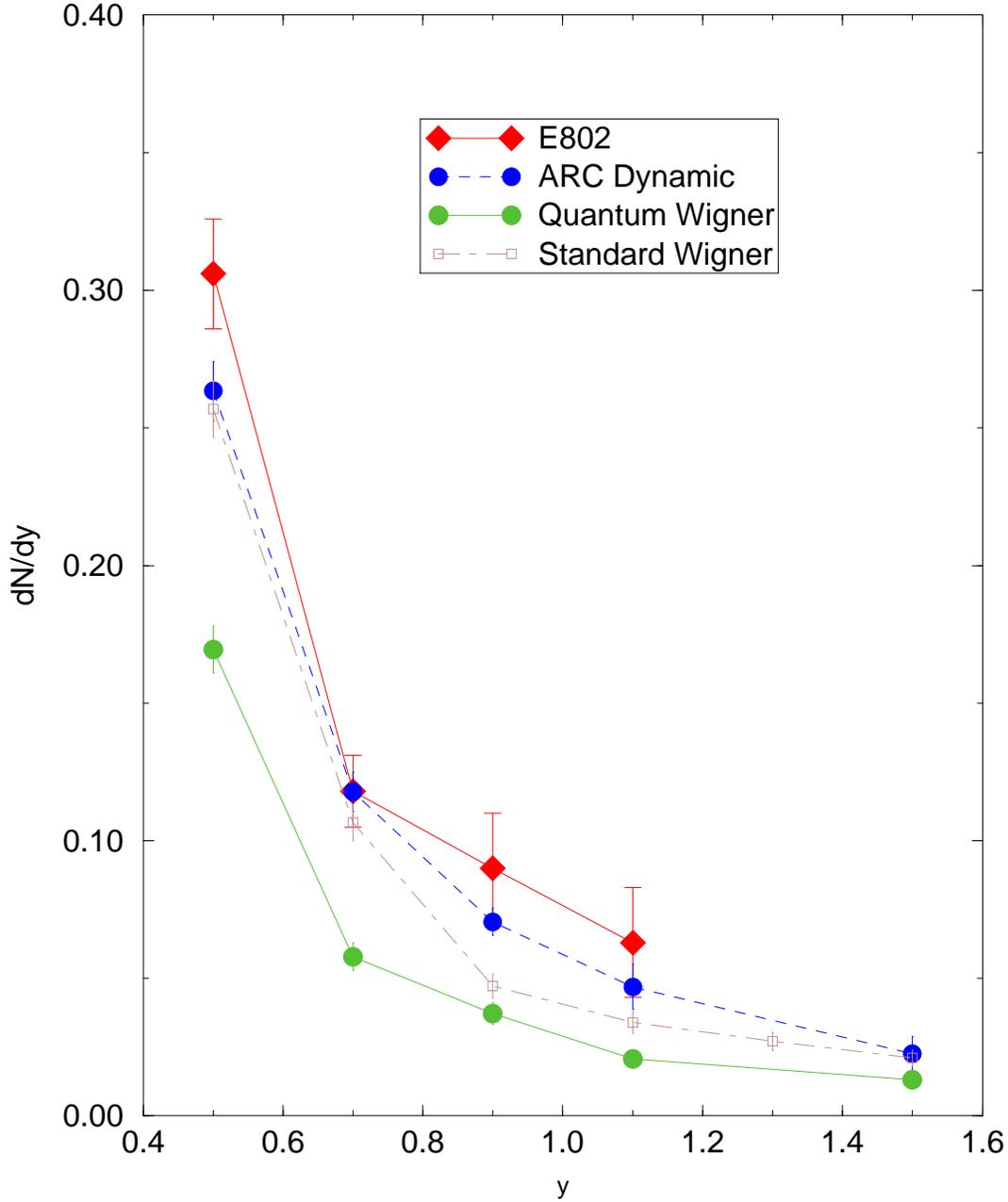}
 \hfil}}
\caption[]{Comparison between ARC Dynamic, Quantum Wigner and
Standard Wigner (obtained using the form factor in Eq.~(16)), but here for the
Si+Au peripheral deuterons. The ratio of peripheral/central yields is close
for ARC Dynamic and Quantum Wigner, but somewhat less for the Standard
Wigner. The latter and ARC also present a changing profile as a function of
rapidity, reflecting contrasting treatments of mainly target-target and
target-projectile coalescences.}
\label{Fig:six}
\end{figure}
\clearpage

\begin{figure}
\vbox{\hbox to\hsize{\hfil
 \epsfxsize=6.4truein\epsffile[24 85 577 736]{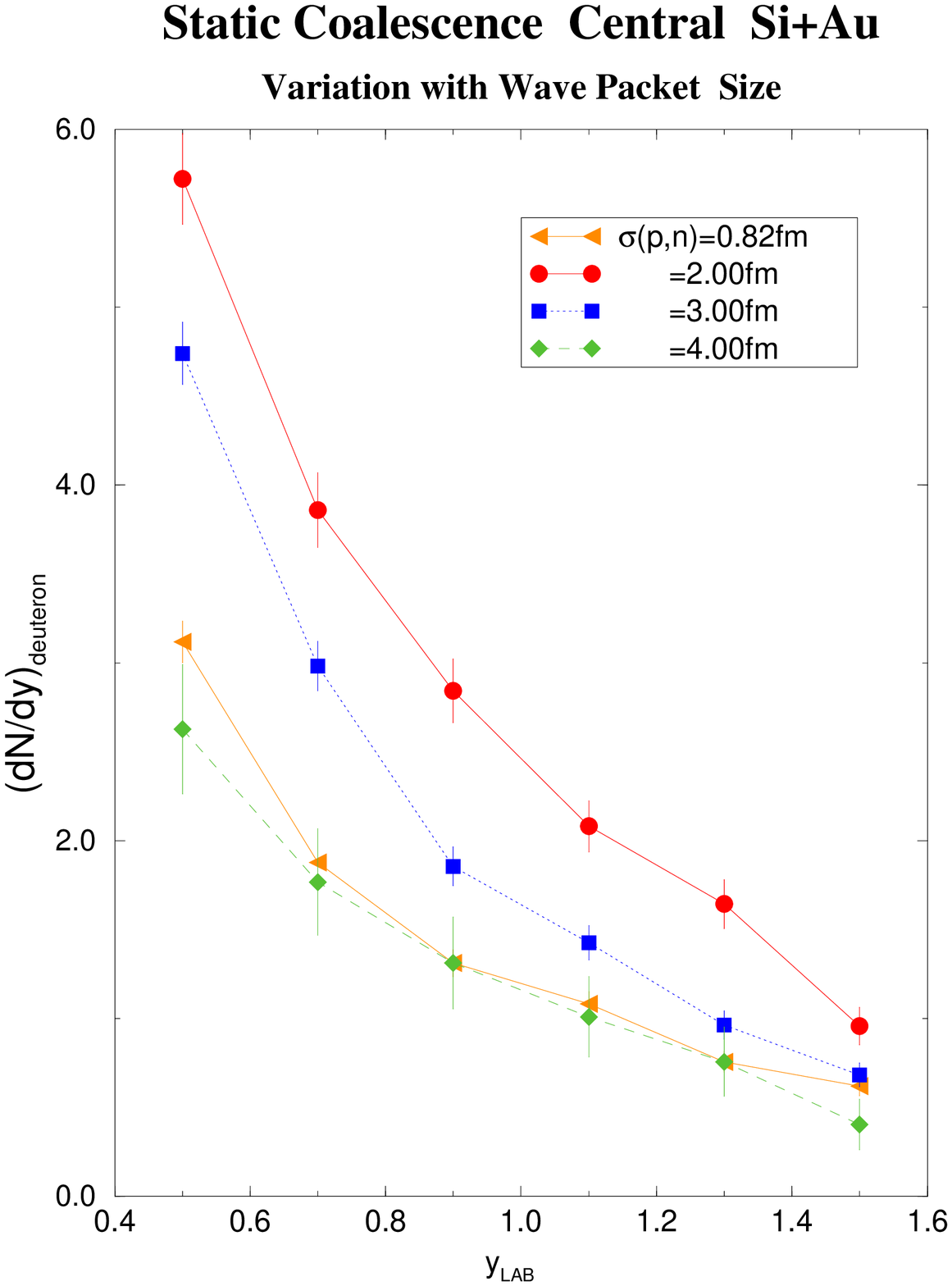}
 \hfil}}
\caption[]{The evolution with wave packet size of the 
ARC Static deuteron central rapidity spectra. There is a 
maximum in magnitude whose position in $\sigma$  depends 
on the folding of the deuteron ``content'' in Eq.~(6) with the ARC 
single neutron distributions in both position and momentum. 
The existence of the maximum is more explicit  in Fig.~9 below.}
\label{Fig:seven}
\end{figure}
\clearpage

\begin{figure}
\vbox{\hbox to\hsize{\hfil
	\epsfxsize=6.4truein\epsffile[24 85 577 736]{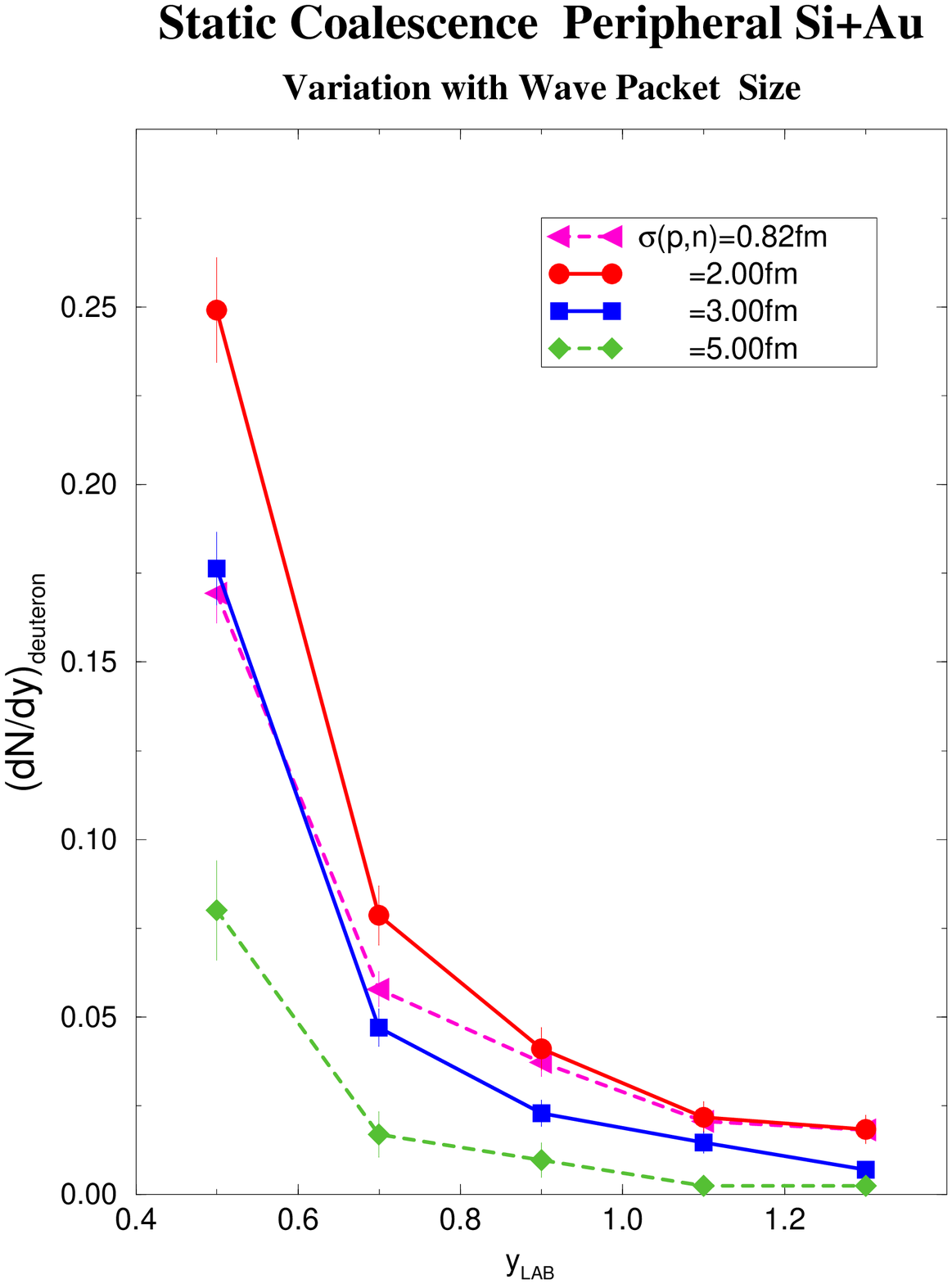}
	\hfil}}
\caption[]{The evolution with wave packet size of the 
ARC Static deuteron peripheral rapidity spectra. See also Fig.~10}
\label{Fig:eight}
\end{figure}
\clearpage

\begin{figure}
\vbox{\hbox to\hsize{\hfil
 \epsfxsize=6.4truein\epsffile[24 85 577 736]{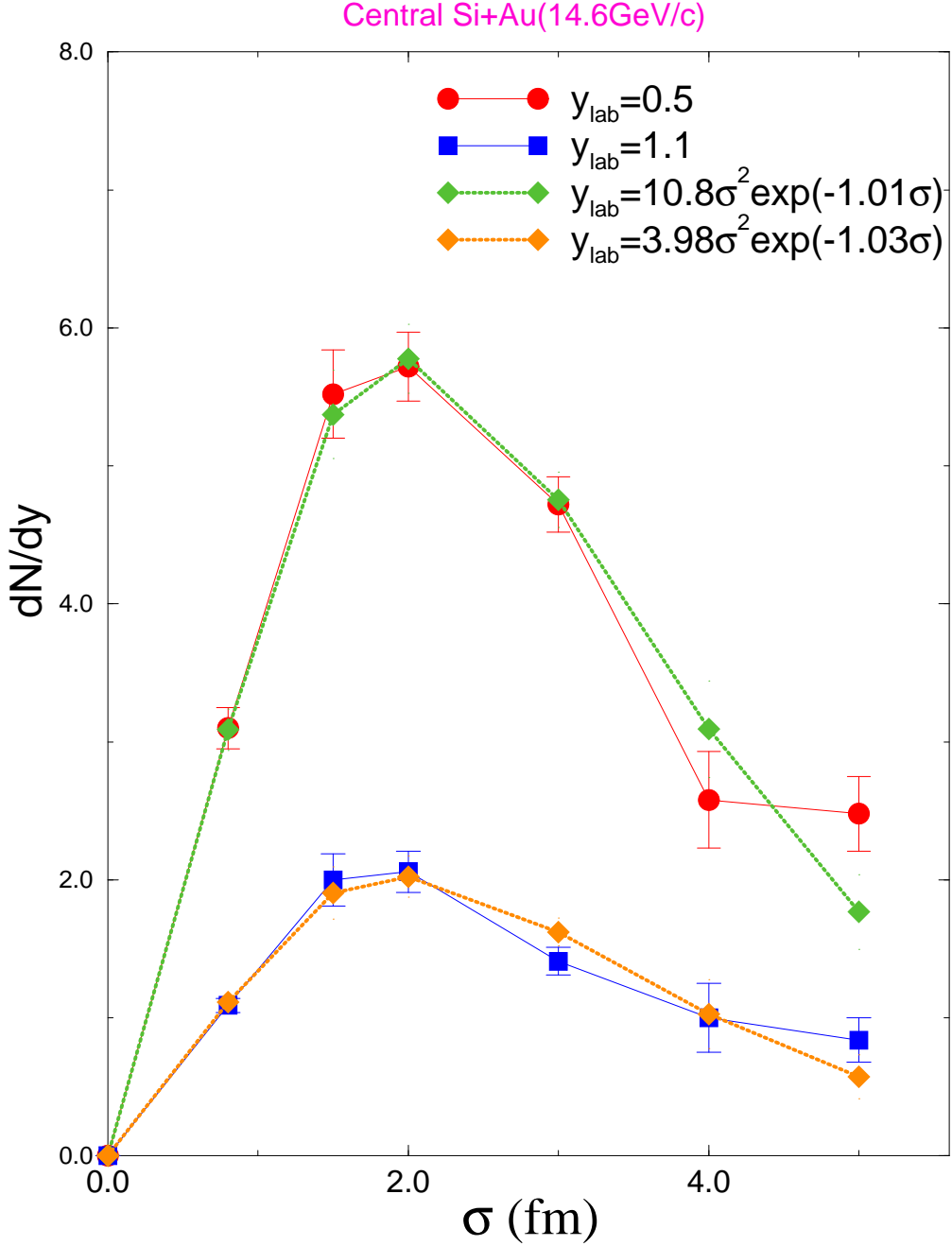}
 \hfil}}
\caption[]{Explicit variation of the central deuteron
$\frac{dN}{dy}$ with size. Eq.~(7) and the associated discussion  suggest the
overlap normalisation is maximized for $\sigma=\frac{\alpha}{\sqrt2}$,
whereas clearly in this figure the maximum occurs nearer to $\sigma=2.0$ for
the differing Static simulations, demonstrating the importance of the wave
packet dynamics. The variation is less marked for the higher, more forward,
rapidity $y=1.1$. Functions fitted to these ARC outputs are also indicated
in this figure, and from these one can in fact extract the position of the
maximum in $\frac{dN}{dy}$ to be close to $\sigma=2$ for central.}
\label{Fig:nine}
\end{figure}
\clearpage

\begin{figure}
\vbox{\hbox to\hsize{\hfil
 \epsfxsize=6.4truein\epsffile[24 85 577 736]{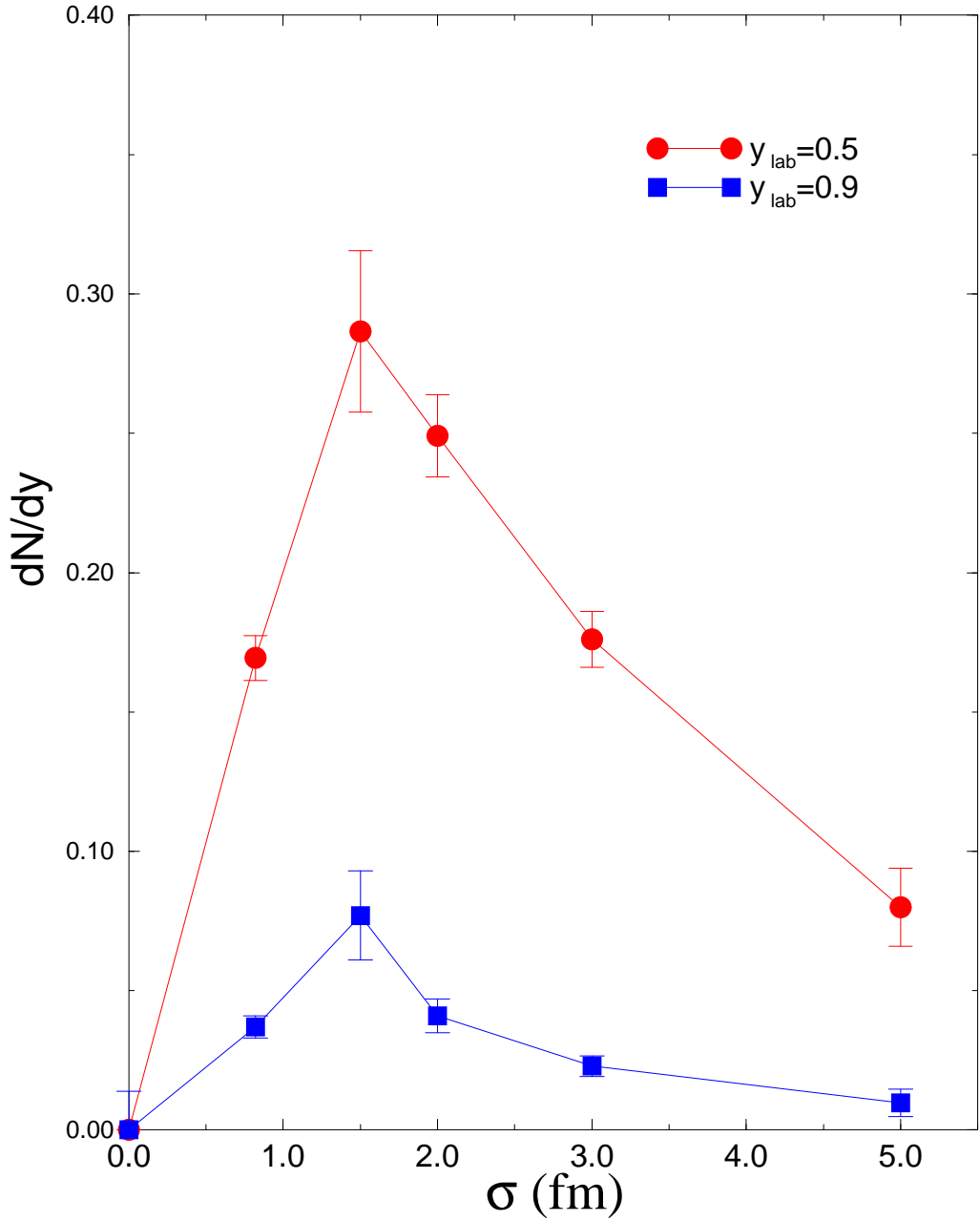}
 \hfil}}
\caption[]{Similar to Fig.~9, but for peripheral 
simulations. The maximum in $\frac{dN}{dy}$ is here below $\sigma=2.$}
\label{Fig:ten}
\end{figure}
\clearpage

\begin{figure}
\vbox{\hbox to\hsize{\hfil
\epsfxsize=6.4truein\epsffile[24 85 577 736]{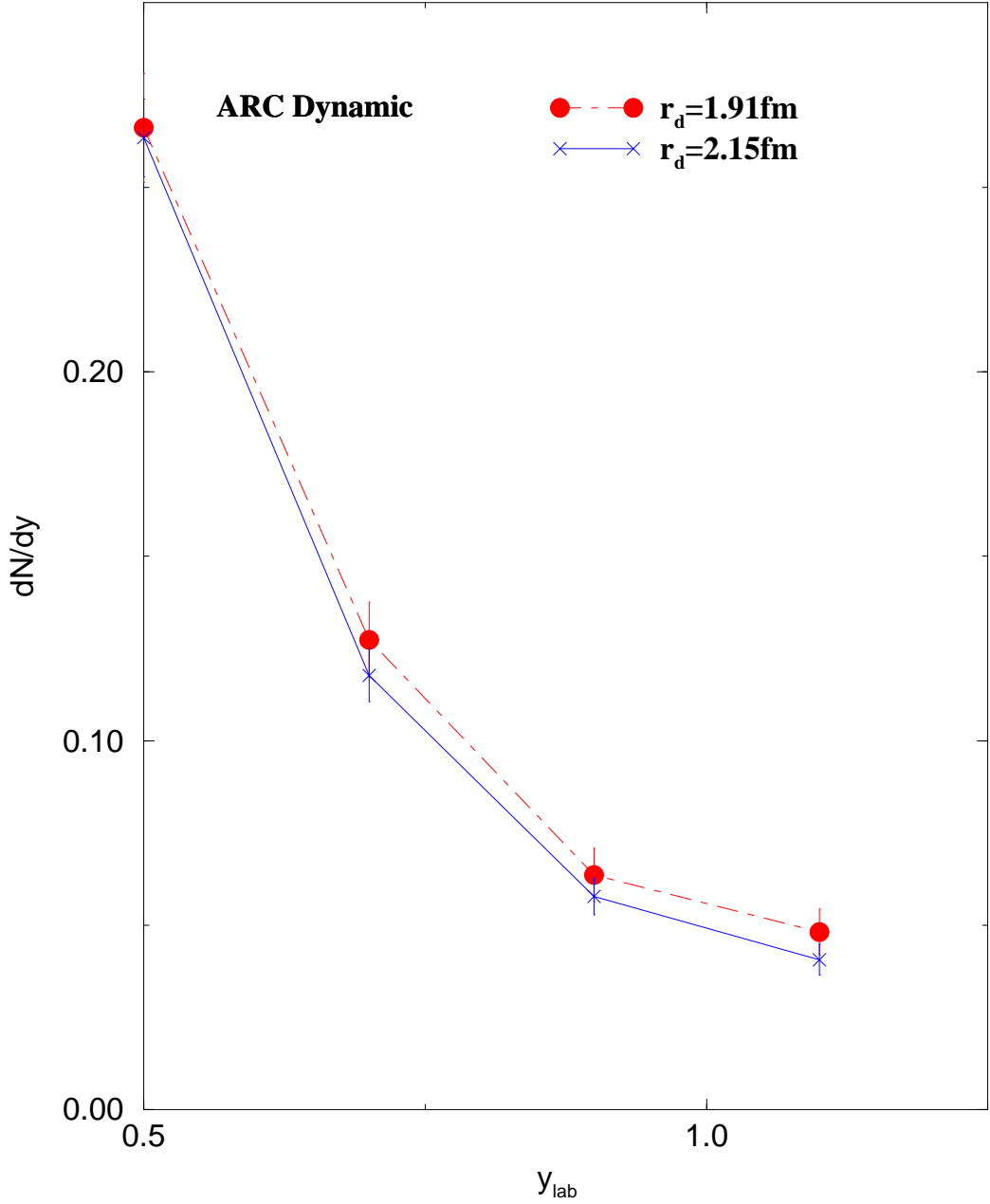}
 \hfil}}
\caption[]{Changes in peripheral rapidity spectra
due to variation in the internal deuteron radius from its point nucleon value
of $1.91fm$ to the charge radius of $2.15fm$ \cite{chargedradii}. Both peripheral
and central rapidity distributions
show a weak dependence on this radius, at least near the actual physical
values for the deuteron size.}
\label{Fig:eleven}
\end{figure}
\clearpage

\begin{figure}
\vbox{\hbox to\hsize{\hfil
\epsfxsize=6.4truein\epsffile[24 85 577 736]{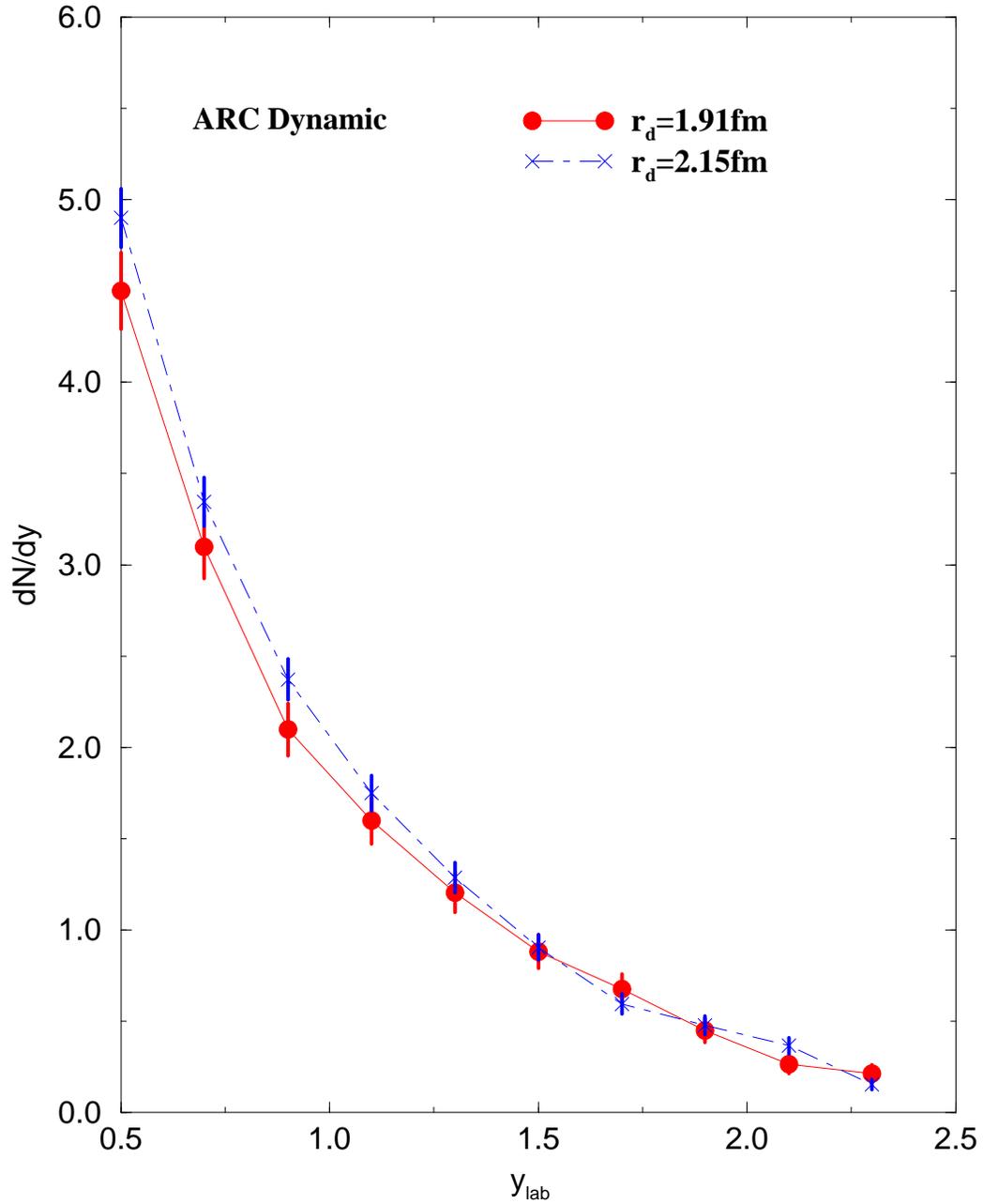}
 \hfil}}
\caption[]{Changes in  central rapidity spectra due
to the above variation in  deuteron radius. Standard
Wigner varies even less with this change in radius.}
\label{Fig:twelve}
\end{figure}
\clearpage

\begin{figure}
\vbox{\hbox to\hsize{\hfil
\epsfxsize=6.1truein\epsffile[24 59 562 736]{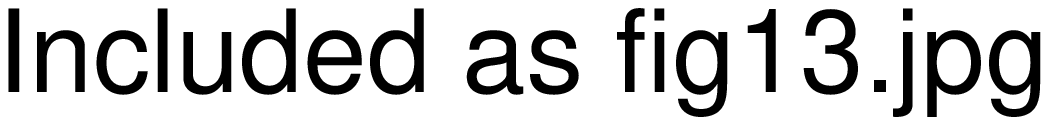}
\hfil}}
\caption[]{Progenitor Pair Sizes vs Rapidity, and Size Histograms: No
rapidity cut.  Wave packet spread in ARC Dynamic is displayed for all np
pairs as well as for only coalesced pairs. In the upper two graphs scatter
plots of all pairs are shown for both central and peripheral collisions: the
three  areas correspond to spectator-spectator (s-s, intermediate
shading), spectator-interacting (s-i, light gray) and interacting-interacting
(i-i, darkest shading). For  peripheral collisions the separation into
target and projectile is clear for the s-s pairs, with the larger sizes
obtaining for the bigger gold nucleus; for central collisions there is no
evidence of s-s pairs.  In the lower two graphs the successful
deuteron-forming pair-distributions are embedded in the overall
histograms. Peripheral coalescence clearly contains (at least) two
components, the smaller sizes correlated to i-i deuteron parentage, the
larger, near $5fm$, to s-i and s-s.}
\label{Fig:thirteen}
\end{figure}
\clearpage

\begin{figure}
\vbox{\hbox to\hsize{\hfil
\epsfxsize=6.1truein\epsffile[24 59 562 736]{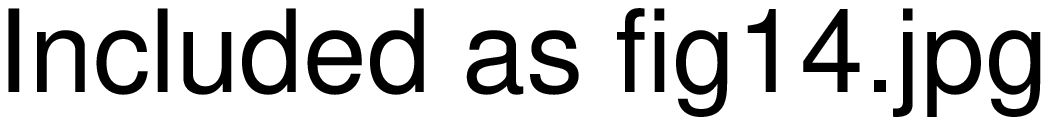}
\hfil}}
\caption[]{Size vs Rapidity and Coalesced Size Distribution:
$y_{lab}>0.5$. For the indicated rapidity cut, which corresponds to the E802
measurements, coalesced pairs for central collisions are almost uniquely from
the s-i and i-i (mid-rapidity only) groups. In the limited events sampled here
deuterons are also formed near projectile rapidities in peripheral
collisions.  The parentage of the successfully coalesced pairs is reflected
in the average wave packet radii, significantly larger for central but
somewhat above the $1fm$ value perhaps expected for peripheral.}
\label{Fig:fourteen}
\end{figure}
\clearpage

\begin{figure}
\vbox{\hbox to\hsize{\hfil
\epsfxsize=6.1truein\epsffile[24 59 562 736]{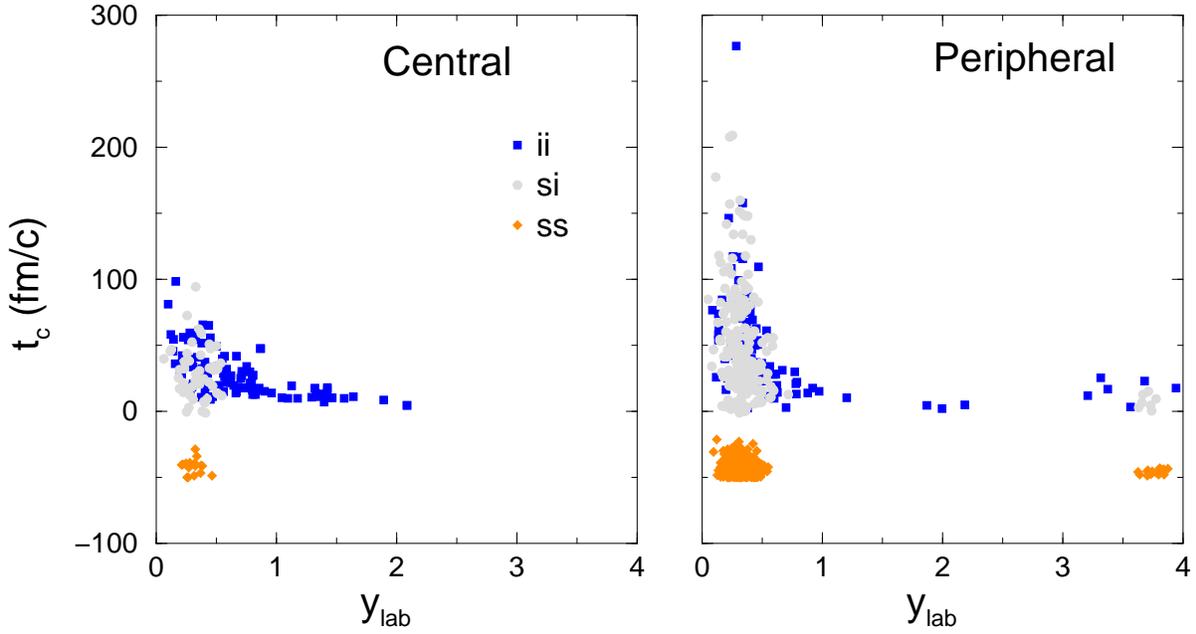}
\hfil}}
\caption[]{Time-Rapidity Structure of Pairs. The time of
coalescence $t_c$, i.e. the last interaction time for either nucleon, in the
cascade global frame is plotted against rapidity for coalesced
deuterons. Clearly in the global frame, i.e. the original equal velocity
frame, the cascade follows interaction and coalescence for appreciable
times. The s-s events, coming earlier and in a narrow time window, are easily
distinguished, while the i-i events spread out appreciably in time.}
\label{Fig:fifteen}
\end{figure}
\clearpage

\begin{figure}
\vbox{\hbox to\hsize{\hfil
\epsfxsize=6.1truein\epsffile[24 59 562 736]{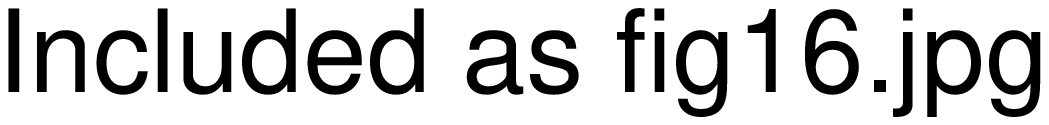}
\hfil}}
\caption[]{Relative Momentum Window vs Rapidity. A most influential parameter for the success of deuteron formation
is the momentum difference between the precursor nucleons. One
notes the large values achieved for the totality of cascading s-i and i-i pairs, the very
small values for all s-s, and contrasts these with the  restrictive,
$\Delta P$, mostly less than 100 MeV/c, for the coalesced pairs. Small
values of $\Delta P$ at coalescence follow from the low energy structure of
the deuteron, and strongly influence the yields as functions of 
peripherality and rapidity. The matching of all pair $\Delta P$'s to the
deuteron wave function passes through the quantum filter in Eq.~7, and 
yields then reflect the overlap dynamics. There is a characteristic rise of
$\Delta P$ with rapidity for the complete set of interacting pairs,
signalling the increase in numbers of interactions towards mid-rapidity.}
\label{Fig:sixteen}
\end{figure}
\clearpage

\begin{figure}
\vbox{\hbox to\hsize{\hfil
\epsfxsize=6.1truein\epsffile[24 59 562 736]{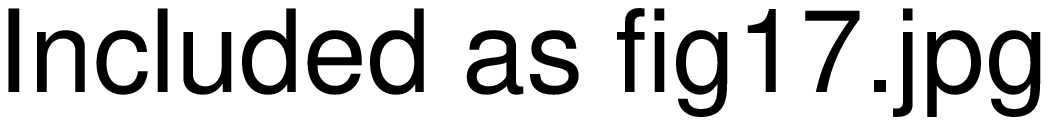}
\hfil}}
\caption[]{Relative Momentum vs Relative Separation. Coalesced deuterons  are displayed in
the lower two graphs, the central collisions to the right. $\Delta X$, the np
separation at coalescence is to be distinguished from the individual nucleon
wave packet size. The coalescence window for this variable is defined by
both the deuteron relative wave function and by the parentage group. The
structure of the overlap factor in Eq.~7 contains a compensation from the
Uncertainty Principle, but some model dependence in overall normalisation
remains. The spread in $\Delta P$ for the spectator-spectator (intermediate
shading) seen in this figure indicate the inclusion of Fermi motion for
target and projectile nucleons.}
\label{Fig:seventeen}
\end{figure}
\clearpage

\begin{figure}
\vbox{\hbox to\hsize{\hfil
\epsfxsize=6.1truein\epsffile[24 59 562 736]{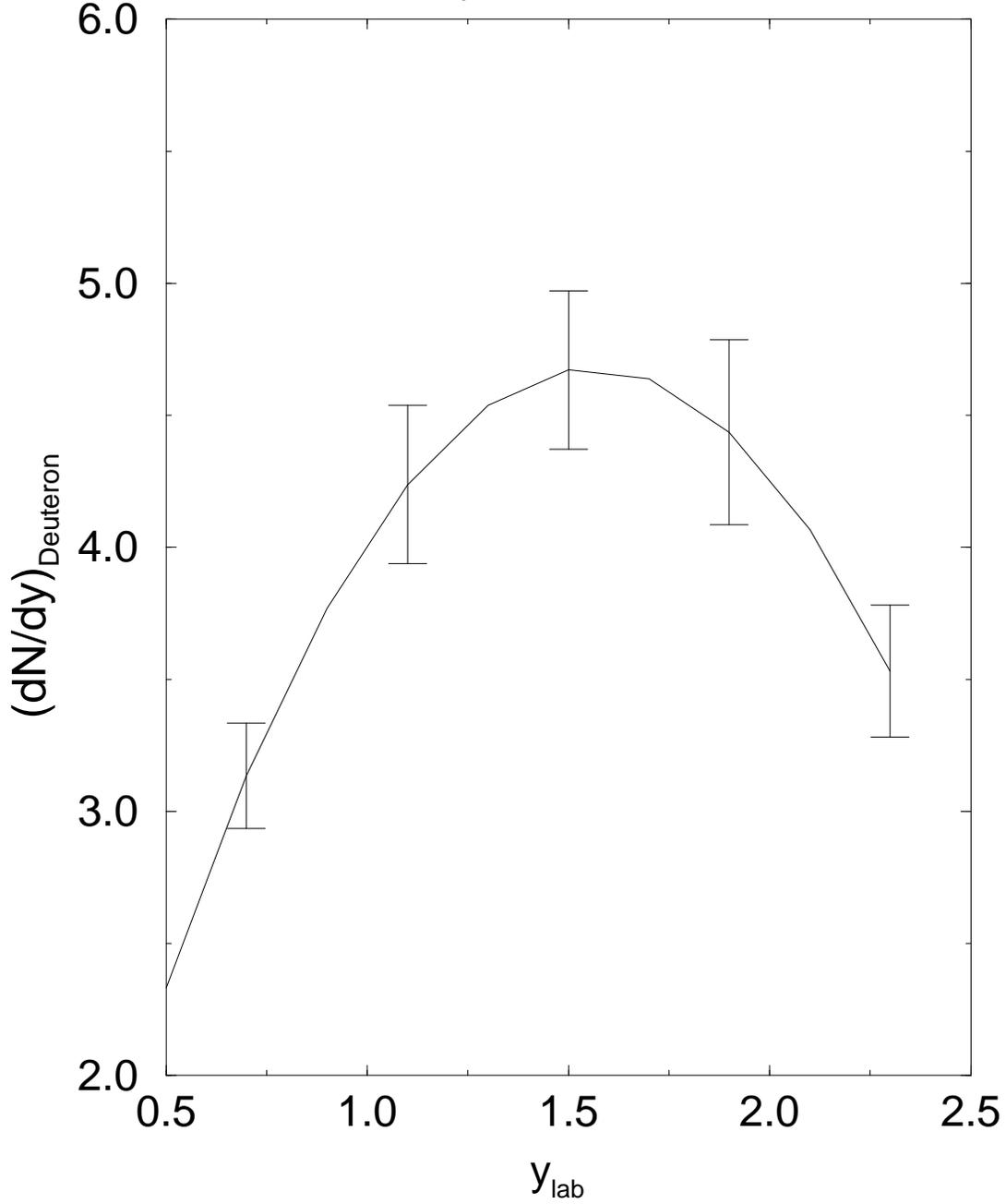}
\hfil}}

\caption[]{\footnotesize Deuterons from ARC Dynamic simulation for Au+Au at
11.6 GeV/c. The centrality cut is defined simply by $b<2$.  Comparison with
forthcoming data must incorporate the actual experimental trigger. The
displayed curve was obtained by fitting a simple parabolic form to the
theoretical data. The latter showed rather more fluctuation but contained
statistical errors of about $10\%$ as indicated.}
\label{Fig:eighteen}
\end{figure}
\clearpage

\begin{figure}
\vbox{\hbox to\hsize{\hfil
 \epsfxsize=6.1truein\epsffile[24 59 562 756]{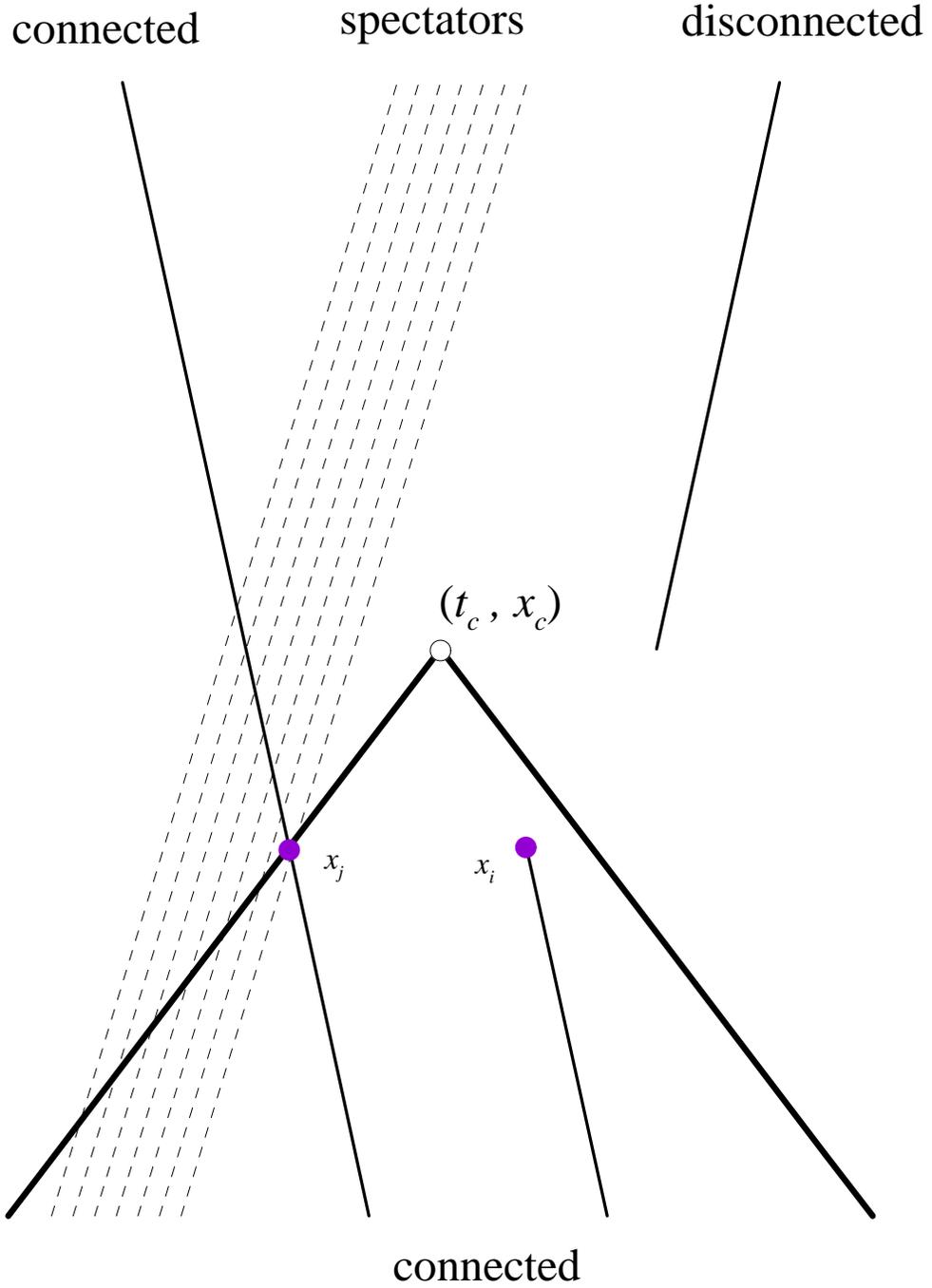}
 \hfil}}
\caption[]{\footnotesize Light-Cone Coalescence.  A
schematic of the space-time picture, at ${\it t}_c$, of particles which enter
into the determination of wave packet size for the nucleon pair potentially
forming a deuteron. The size parameter $\sigma$ is obtained by averaging over
the positions of all pair comovers, as defined in the text.}
\label{Fig:nineteen}
\end{figure}
\clearpage

\end{document}